\documentclass[aps,prd,nofootinbib,twocolumn,superscriptaddress,preprintnumbers,balancelastpage,longbibliography]{revtex4-1}

\usepackage{appendix, amsmath, amssymb}
\usepackage[english]{babel}
\usepackage{bm, booktabs}
\usepackage{color}
\usepackage{float, fontawesome5, filecontents}
\usepackage{graphicx}
\graphicspath{{figures/}}
\usepackage{hepunits}
\usepackage[utf8]{inputenc}
\usepackage{mathtools, multirow}
\usepackage{physics, pifont}
\usepackage{soul}
\usepackage{tensor}
\usepackage{url}
\usepackage[dvipsnames]{xcolor}
\usepackage{xspace}

\usepackage{hyperref}
\hypersetup{
    colorlinks=true,       
    linkcolor=blue,        
    citecolor=blue,        
    filecolor=magenta,     
    urlcolor=blue          
}

\newcommand{\ra}[1]{\renewcommand{\arraystretch}{#1}}

\makeatletter
\def\hlinewd#1{%
\noalign{\ifnum0=`}\fi\hrule \@height #1 \futurelet
\reserved@a\@xhline}
\makeatother
\newcolumntype{L}[1]{>{\raggedright\let\newline\\\arraybackslash\hspace{0pt}}m{#1}}
\newcolumntype{C}[1]{>{\centering\let\newline\\\arraybackslash\hspace{0pt}}m{#1}}
\newcolumntype{R}[1]{>{\raggedleft\let\newline\\\arraybackslash\hspace{0pt}}m{#1}}
\renewcommand{\arraystretch}{1.2}

\newcommand{\expn}[1]{\left\langle#1\right\rangle}

\newcommand{\xray}{\textit{X}-ray\xspace}

\newcommand{\swave}{$s$-wave\xspace}
\newcommand{\pwave}{$p$-wave\xspace}

\newcommand{\cmfast}{\texttt{21cmFAST}\xspace}
\newcommand{\dmcm}{\texttt{DM21cm}\xspace}
\newcommand{\dhis}{\texttt{DarkHistory}\xspace}
\newcommand{\bh}{\texttt{BlackHawk}\xspace}

\newcommand{\Mpbh}{M_\text{PBH}}
\newcommand{\Mdotpbh}{\dot{M}_\text{PBH}}
\newcommand{\Msun}{M_\odot}

\newcommand{\cin}{c_\text{in}}

\newcommand{\PR}{\textbf{PR}\xspace}
\newcommand{\BHL}{\textbf{BHL}\xspace}
\newcommand{\BHLmt}{\textbf{BHL-}$M_\text{th}^\text{low}$\xspace}
\newcommand{\PRcm}{\textbf{PR-}$\cin^-$\xspace}
\newcommand{\PRcp}{\textbf{PR-}$\cin^+$\xspace}
\newcommand{\PRB}{\textbf{PR-}$B$\xspace}
\newcommand{\PRH}{\textbf{PR-}$H$\xspace}

\newcommand{\PRdp}{\textbf{PR-}$\delta_e^+$\xspace}
\newcommand{\PRdm}{\textbf{PR-}$\delta_e^-$\xspace}

\begin{document}

\title{Constraining inhomogeneous energy injection from annihilating dark matter and primordial black holes with 21-cm cosmology}

\author{Yitian Sun}
\email{yitian.sun@mcgill.ca}
\affiliation{Department of Physics, McGill University, Montreal, QC H3A 2T8, Canada}
\affiliation{Trottier Space Institute at McGill, Montreal, QC H3A 2T8, Canada}

\author{Joshua W. Foster}
\affiliation{Department of Physics, University of Wisconsin-Madison, Madison, WI 53706, USA}

\author{Julian B.~Mu\~{n}oz}
\affiliation{Department of Astronomy, The University of Texas at Austin, 2515 Speedway, Stop C1400, Austin, TX 78712, USA}
\affiliation{Cosmic Frontier Center, The University of Texas at Austin, Austin, TX 78712}

\date{\today}

\begin{abstract}
The cosmic dawn 21-cm signal is a highly sensitive probe of any process which injects energy into the intergalactic medium, enabling novel searches for anomalous energy injection by through dark matter interactions. In addition to modifying the global 21-cm signal, these processes would leave distinct imprints on the frequency-resolved 21-cm power spectrum with a morphology jointly set by the time- and spatial-dependence of energy emission and absorption. In this work, we extend the \texttt{DM21cm} code package, which models the effects of spatially inhomogeneous energy emission and deposition on 21-cm cosmology, to study three well-motivated scenarios which 21-cm measurements are particularly well-suited to probe: dark matter annihilation through \textit{p}-wave processes, the Hawking radiation of light primordial black holes, and energetic emission from accreting solar-mass primordial black holes. We project sensitivities to each of these scenarios, demonstrating that leading or near leading sensitivity to each can be achieved through 21-cm probes.  We also make public our updates to the \texttt{DM21cm} code package that enable it to accommodate energy injection processes for general continuum spectra with arbitrary spatial and temporal dependence in an accompanying release \href{https://github.com/yitiansun/DM21cm}{\faGithub}.
\end{abstract}
\maketitle

\section{Introduction}

Observations of the redshifted 21-cm line produced by the hyperfine transition of neutral hydrogen provide a unique opportunity to probe cosmology at intermediate redshifts at times between recombination and reionization. In particular, the 21-cm emission that traces the three-dimensional distribution of neutral hydrogen enables a clean probe of fundamental cosmology, the thermal and ionization history of our universe, and the early stages of structure formation \cite{Furlanetto:2006jb,Pritchard2012,Mondal:2023xjx}. Ongoing and upcoming 21-cm measurements include: EDGES~\cite{Monsalve:2016xbk}, LEDA~\cite{Spinelli:2022xra}, PRI$^\mathcal{Z}$M~\cite{prizm}, SARAS~\cite{Singh:2017syr}, SCI-HI~\cite{Voytek:2013nua, Peterson:2014rga}, BIGHORNS~\cite{Sokolowski_2015}, and REACH~\cite{deLeraAcedo:2022kiu}, which measure the global 21-cm monopole signal; and radio interferometers like PAPER~\cite{Pober:2013ig}, the MWA~\cite{Tingay:2012qe}, LOFAR~\cite{Rottgering:2003jh}, HERA~\cite{DeBoer:2016tnn}, and SKA~\cite{Dewdney:2009tmd}, which measure the spatially varying on-sky 21-cm emission, often summarized in terms of a 21-cm power spectrum, though higher-point functions and nongaussianities may also be of interest~\cite{Munoz:2015eqa}.

The local intensity of 21-cm line emission is jointly determined by the temperature and ionization of the intergalactic medium (IGM), and, in $\Lambda$CDM cosmology, heating and ionization of the IGM are expected to be driven by energy injection from stellar processes. As a result, the $\Lambda$CDM prediction for the 21-cm signal is largely determined by the local star formation rate (SFR), which becomes appreciable at $z\sim10-20$ when the first star-hosting galaxies form  \cite{Bromm:2003vv}. The absence of efficient astrophysical mechanisms for energy injection at earlier times then makes 21-cm measurements particularly sensitive to energy injection from dark matter. This includes processes that source energetic, electromagnetically interacting particles that then deposit their energy in the IGM. Moreover, even when anomalous and standard astrophysical energy injection occur simultaneously, differences in the spatial morphology of their imprints on the 21-cm may still enable the identification of new physics in the 21-cm power spectrum. 

\begin{figure}[!ht]
    \centering
    \includegraphics[width=\linewidth]{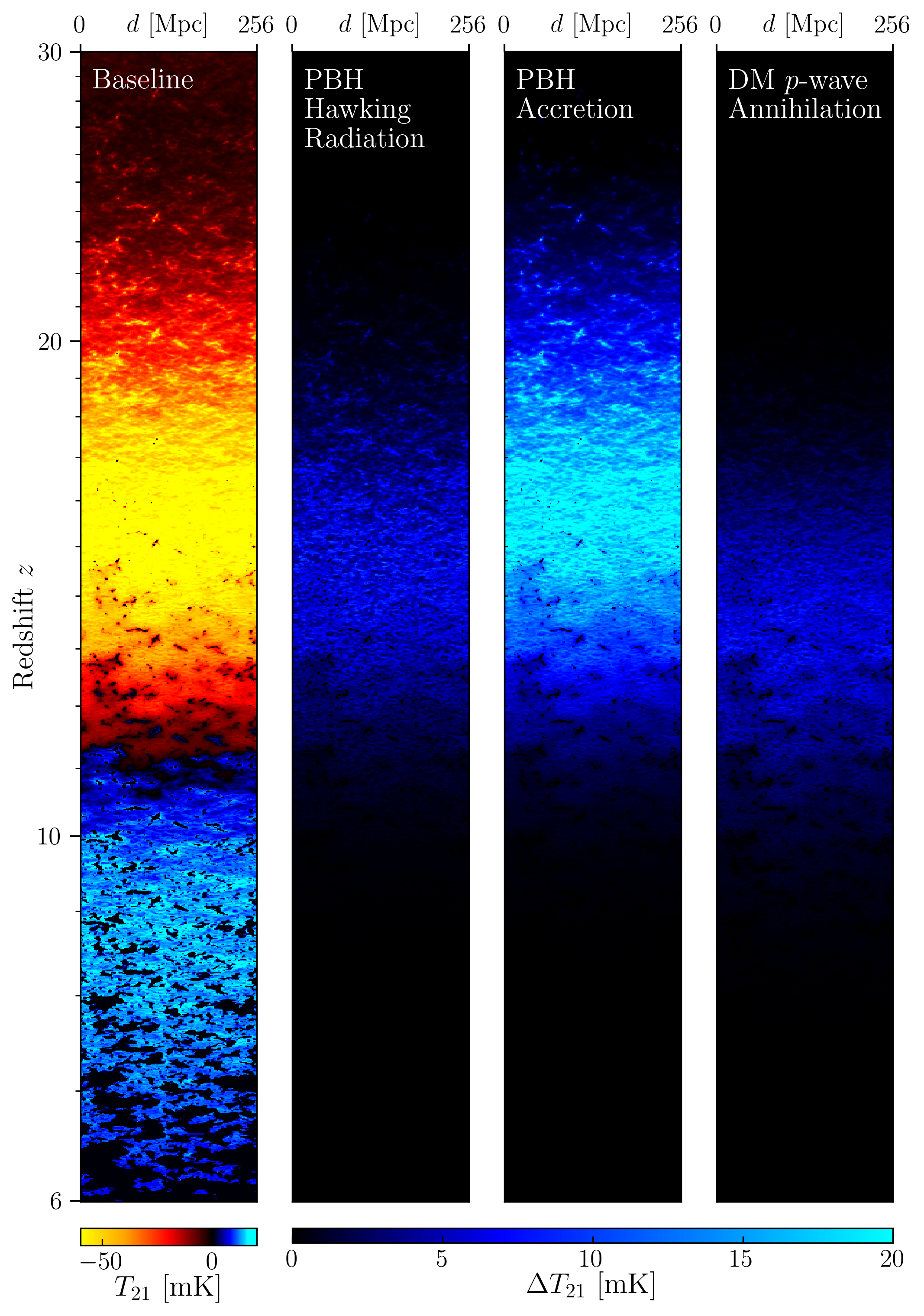}
    \caption{\textbf{$T_{21}$ lightcones for various energy injection scenarios.} The first panel shows a baseline $T_{21}$ with fiducial astrophysical parameters and no exotic energy injections. The right three panels show changes to the $T_{21}$ lightcone under the scenarios of i. DM \pwave annihilation to photons ($m_\chi=100$~keV); ii. PBH Hawking radiation ($\Mpbh=10^{16.5}$~g); iii. PBH accretion ($\Mpbh=10^4\Msun$). In each example, the strength of the exotic energy injection is taken to be at the projected 95$^\mathrm{th}$ percentile upper limit under the HERA sensitivity developed in this work.}
    \label{fig:illustrative_example}
\end{figure}

Measurements of the 21-cm signal power spectrum that can leverage the distinct spatial and spectral morphology of a DM-induced signal are particularly compelling for probing exotic energy injection via DM processes. On the other hand, developing predictions for the 21-cm signal in these contexts can be challenging, since, at the redshifts relevant for producing 21-cm radiation, the structure formation process has generated $\mathcal{O}(1)$ overdensities on cosmologically relevant (Mpc) scales. The impact of these inhomogeneities --- both on the spatial distribution of DM processes that generate energetic radiation and on the efficiency with which that radiation heats and ionizes the IGM --- must be carefully accounted for in order to make robust predictions of the associated 21-cm signal. In addition, one must include the backreaction whereby modifications to the IGM induced by these processes further alter the rates of energy deposition.

In previous work, we developed a new code, \texttt{DM21cm}, which models the effect of spatially inhomogeneous energy injection on the 21-cm signal by interfacing the semi-numerical \texttt{21cmFAST} simulation framework~\cite{Mesinger_2010} with the modeling of IGM heating, ionization, and excitation by highly energetic cosmic rays with the \texttt{DarkHistory} code package~\cite{Liu:2019bbm, Sun:2022djj, Liu:2023fgu,Liu:2023nct}. Initial studies have demonstrated that 21-cm power spectrum measurements anticipated from HERA have the potential to set leading astrophysical constraints on the  decays of $\sim$keV (MeV) DM to monoenergetic photons (electrons)~\cite{Sun:2023acy}. While already physically interesting, these scenarios represent only a limited subset of scenarios in which energy injection by DM processes can lead to detectable signatures in the 21-cm power spectrum. Moreover, monochromatic decays represent the energy injection most straightforwardly accommodated within the \texttt{DM21cm}, while alternative new-physics scenarios may prescribe more general time- and spatial-dependence in the rate and spectrum of DM-induced emission. 

In this work, we apply \texttt{DM21cm} to study 21-cm sensitivities to three motivated new physics scenarios that realize more complicated energy injection histories than that of monochromatic decays. Those scenarios are: Hawking radiation from evaporating light primordial black holes (PBHs), cosmic ray emission from heavier accreting PBHs, and \textit{p}-wave annihilation of DM. In the case of evaporating light PBHs, while the PBH emission realizes a decay-like spatial morphology like that previously studied in~\cite{Sun:2023acy}, the Hawking radiation spectrum is a continuum and undergoes relevant time-evolution in scenarios where the PBHs lose an appreciable amount of their mass. In the case of PBH accretion, the accretion rate and the luminosity of associated emission is affected by the density and temperature of baryonic matter, as well as its relative velocity with respect to the PBH, endowing it with considerable environmental dependence. Similarly, the rate of \textit{p}-wave annihilation is jointly set by the DM density and velocity dispersion, making it sensitive to the times and locations at which large DM halos form. Moreover, while each of these scenarios can also be probed with Cosmic Microwave Background (CMB) measurements, they are more naturally studied in the context of 21-cm cosmology as the majority of their energy injection occurs at later times than would most sensitively probed by the CMB.

As an illustrative example of these three scenarios, in Fig.~\ref{fig:illustrative_example}, we show the changes in $T_{21}$ lightcones from a fiducial \dmcm simulation under the (i) Hawking radiation of $10^{16.5}$~g PBHs, initially making up $f_\text{PBH}=1.06\times10^{-4}$ of DM; (ii) accretion emission of $10^4~\Msun$ PBHs making up $f_\text{PBH}=2.24\times10^{-5}$ of DM; and (iii) \pwave annihilation of a 100~keV DM particle to two photon with a cross section of $C_\sigma=5.72\times10^{-21}$~cm$^3$/s. These injection strengths correspond to the projected $95^\text{th}$ percentile upper limit of HERA sensitivity used in this work, accounting for uncertainties on astrophysical quantities. The difference in their injection behavior and the resulting effect on sensitivity can be clearly seen.

This paper is organized as follows. In Sec.~\ref{sec:review}, we review basic aspects of 21-cm cosmology with an emphasis on the modeling of inhomogeneous anomalous energy injection on the 21-cm power spectrum performed with \texttt{DM21cm}. In Sec.~\ref{sec:pbh_evaporation}, Sec.~\ref{sec:pbh_accretion}, and Sec.~\ref{sec:dm_annihilation}, we review, describe the \texttt{DM21cm} implementation, and project HERA sensitivities to the evaporating PBH, accreting PBH, and \textit{p}-wave annihilating DM scenarios, respectively. Finally, we provide some concluding remarks in Sec.~\ref{sec:conclusion}.

\section{Summary of 21-cm Cosmology and the Modeling Pipeline}
\label{sec:review}
In this section, we begin with a brief review of the relevant aspects of 21-cm cosmology and our implementation of exotic energy injection in \texttt{DM21cm}. For a more thorough review of 21-cm cosmology, we refer readers to, \textit{e.g.}, \cite{2012RPPh...75h6901P}, while for a more detailed description of \texttt{DM21cm} code, we refer readers to \cite{Sun:2023acy}. 

\subsection{Review of 21-cm Cosmology and \texttt{21cmFAST}}

The fundamental observable of 21-cm cosmology is the brightness temperature of the redshifted 21-cm line relative to the CMB black body temperature. Along the line-of-sight, this is a frequency-dependent quantity given by
\begin{equation}
\begin{split}
    T_{21}(\nu) &\approx 27 x_\mathrm{HI}(1+\delta)\left( \frac{H}{dv_r/dr + H}\right)\left( 1-\frac{T_\gamma}{T_S}\right) \\
    &\times \left(\frac{1+z}{10} \frac{0.15}{\Omega_M h^2} \right)^{1/2} \left( \frac{\Omega_b h^2}{0.023} \right)\,\mathrm{mK},
\end{split}
\end{equation}
where $\delta$ is the Eulerian density contrast, $H$ is the Hubble parameter, and $dv_r/dr$ is the comoving gradient of the comoving velocity projected along the line of sight. $T_\gamma$ is the CMB temperature, $\Omega_M$ and $\Omega_b$ the respective present-day matter and baryon abundances relative to the critical density, and $h$ the present-day Hubble parameter in units of $100\,\mathrm{km/s/Mpc}$ \cite{Furlanetto:2006tf}. While each of these depends on the cosmology, they are otherwise independent of any exotic energy injection, which may leave imprints on the neutral hydrogen fraction $x_\mathrm{HI}$ and the gas spin temperature $T_S$.

\subsubsection{Spin Temperature Evolution}
The gas spin temperature is jointly determined by three effects: the absorption and emission of CMB photons; collisions of hydrogen atoms with other hydrogen atoms, free electrons, and free protons; and the absorption and emission of Lyman-$\alpha$ photons through the Wouthuysen-Field coupling. In total, $T_S$ is given 
\begin{equation}
        T_S^{-1} = \frac{T_\gamma^{-1} + x_c T_k^{-1} + x_\alpha T_\alpha^{-1}}{1 + x_c + x_\alpha},
\end{equation}
where $T_k$ is the gas kinetic temperature and $T_\alpha$ is the effective color temperature of the Ly$\alpha$ radiation field; $x_c$ and $x_\alpha$ are coupling coefficients for the collision and Ly$\alpha$ scatterings~\cite{Furlanetto:2006jb}. The color temperature is well-approximated by the gas kinetic temperature~\cite{1959ApJ...129..536F, 2006MNRAS.367..259H}, and $x_c$ has been calculated in detail as a function of the kinetic temperature in \cite{2005ApJ...622.1356Z, Furlanetto:2006su}. The role of energy injection, exotic or otherwise, is then to drive a time-evolving kinetic temperature $T_k$ and to modify the Wouthuysen-Field coupling. 

Following the \texttt{21cmFAST} modeling, the kinetic temperature evolution is described by
\begin{equation}
\begin{split}
\label{Eq:ODE}
\frac{d x_e(z, \mathbf{x})}{dz} &= \frac{dx_e^\mathrm{exotic}}{dz}+\frac{dt}{dz} \left[\Lambda_\mathrm{ion} - \alpha_A C x^2_e n_A f_\mathrm{H} \right] \\
\frac{dT_k(z, \mathbf{x})}{dz} &= \frac{dT_k^\mathrm{exotic}}{dz} + \frac{2}{3k_B(1+x_e)} \frac{dt}{dz} \sum_p \epsilon_p\\
&+ \frac{2 T_k}{3 n_A}\frac{dn_A}{dz} - \frac{T_k}{1+x_e} \frac{dx_e}{dz}.
\end{split}
\end{equation}
where $x_e$ is the local ionized fraction in the ``mostly neutral" IGM generated by photoionizing cosmic rays, $n_A$ the local physical nuclear density, $\alpha_A$ the case-A recombination coefficient, $C\equiv\expn{n_e^2}/\expn{n_e}^2$
the free-electron clumping factor, $k_B$ the Boltzmann constant, and $f_\text{H}\equiv n_\text{H}/(n_\text{H}+n_\text{He})$ the hydrogen nucleus number fraction. The $\epsilon_p$ and $\Lambda_\mathrm{ion}$  are, respectively, the local heating rate (summed over processes $p$) and the local ionization rate associated with photoionizing \textit{X}-rays generated through standard astrophysical, \textit{i.e.}, stellar, processes. The Wouthuysen-Field coupling is also minimally modified in the presence of exotic energy injection, taking the form
\begin{equation}
    x_\alpha = 1.7 \times 10^{11} \frac{S_\alpha}{1+z} \left(\frac{J_\alpha + J_\alpha^\mathrm{exotic}}{\mathrm{cm}^2 \,\mathrm{s}^{-1} \, \mathrm{Hz}^{-1} \,\mathrm{sr}^{-1}} \right)
\end{equation}
where $S_\alpha$ is an atomic physics correction factor \cite{2006MNRAS.367..259H}, $J_\alpha$ is the spatially dependent Ly$\alpha$ intensity generated through standard astrophysical processes, and  $J_\alpha^\mathrm{exotic}$ is the spatially dependent Ly$\alpha$ intensity generated through exotic energy-injecting processes.

The rates and intensities associated with the standard astrophysical processes are implemented in \cmfast with a formalism described in \cite{Mesinger_2010} and expanded upon in \cite{Park:2018ljd, Qin:2020xyh, Munoz:2021psm}. We do not detail them at greater length here.

\subsubsection{Neutral Hydrogen Evolution}
In \cmfast, the ionization fraction of hydrogen is jointly driven by the weakly ionizing \textit{X}-ray emission and strongly ionizing UV emission. The effect of the weakly ionizing \textit{X}-ray emission is captured through the ``mostly neutral" ionization fraction $x_e$ evolved as part of Eq.~\ref{Eq:ODE} while the fraction of hydrogen which has not been ionized by UV emission is calculated via a filter-based excursion-set formalism \cite{Park:2018ljd,Mesinger:2007pd} and is denoted by $\mathrm{x}_\mathrm{HI}^\text{filter}$ .

In total, the neutral hydrogen fraction is given by
\begin{equation}
    x_\mathrm{HI} = \mathrm{max}[0, x_\mathrm{HI}^\mathrm{filter} - x_e].
\end{equation}
Since the ionizing effect of exotic energy injection considered in this work is fully captured by the modified evolution of $x_e$, we then realize an accurate and self-consistent evolution of $x_\mathrm{HI}$.

\subsection{21-cm Modeling in \texttt{DM21cm}}
\label{sec:ExoticImplementation}
The \texttt{DM21cm} code package extends the \cmfast evaluation to account for exotic energy injection and deposition by evolving the distribution and spectrum of energetic photons beginning from the time at which they are generated through an exotic process of interest. We consider processes that locally inject a spectrum of photons and  electrons with a volumetric rate
\begin{equation}
    \frac{dN_{\gamma/e}}{dE dV dt}(z, \mathbf{x})= \frac{dN_\text{inj}}{dVdt}(z, \mathbf{x})\times \frac{dN_{\gamma/e}}{dE}(z)
    \label{eq:EnergyInjectionRate}
\end{equation}
where $dN_\text{inj}/dVdt$ is the rate of injection event per unit volume, $dN_{\gamma/e}/dE$ is the spectrum of secondary photons and electrons produced by a single injection event, $z$ is the redshift, and $\mathbf{x}$ is the location, allowing for dependence on, \textit{e.g.}, the local overdensity parameter $\delta(\mathbf{x})$. This prescription generalizes the scenarios studied in \cite{Sun:2023acy}, which considered only monochromatic decays of dark matter of mass $m_\chi$ and lifetime $\tau$ so that $dN/dE = 2 \delta(E-m_\chi/2)$ and $dN_\text{inj}/dVdt = \bar\rho(1 +\delta) /m_\chi \tau$. 

In \texttt{DM21cm}, the energy deposition by these injected is implemented using transfer functions developed from \dhis. We summarize the salient details of energy deposition from these injected electrons and photons, but see \cite{Sun:2023acy} for a full characterization.

\subsubsection{Energy Deposition from Electrons}
Electrons typically deposit their energy in a manner that is instantaneous and on the spot, and they are treated with two transfer functions. The first is $T_{\gamma e}(\delta, x_\mathrm{HI}, z, \Delta z)$, which relates an input spectrum of electrons per baryon $dN_e^\mathrm{in}/dE$ to an outgoing spectrum of photons produced by scattering processes involving those electrons at their scattering products which occur at times between $z$ and $z + \Delta z$ and at a local overdensity $\delta$ and local neutral hydrogen fraction $x_\mathrm{HI}$. 
\begin{equation}
\begin{gathered}
    \frac{dN_\gamma^\text{out}}{dE} = T_{\gamma e}(\delta, x_\mathrm{HI} | z, \Delta z) \frac{dN_e^\text{in}}{dE}.
\end{gathered}
\end{equation}
The second transfer function is $D_{ce}(\delta, x_\mathrm{HI} | z, \Delta z)$, which relates an input electron spectrum to the change in the kinetic temperature $\Delta T_k$, the change in the ionization fraction $\Delta x_e$, and the Ly$\alpha$ intensity $J_\alpha$ their scatterings produce at times between $z$ and $z + \Delta z$ and at a local overdensity $\delta$ and local neutral hydrogen fraction $x_\mathrm{HI}$. In summary, 
\begin{equation} \label{eq:electronTF}
\begin{gathered}
    \frac{dN_\gamma^\text{out}}{dE} = T_{\gamma e}(\delta, x_\mathrm{HI} | z, \Delta z) \frac{dN_e^\text{in}}{dE} \\
    \begin{bmatrix} \Delta T_k \\ \Delta x_e \\ J_\alpha \end{bmatrix} = D_{ce}(\delta, x_\mathrm{HI} | z, \Delta z) \frac{dN_e^\text{in}}{dE},
\end{gathered}
\end{equation}
where $dN_\gamma^\text{out} / dE$ is the spectrum of outgoing photons per baryon.
 
At each timestep of the \dmcm evaluation, we include these contributions in the time-evolution of the kinetic temperature, ionization fraction, and spin temperature performed by \cmfast. The spectrum of outgoing photons is then cached and treated as a local photon injection spectrum, which we describe subsequently.

\subsubsection{Energy Deposition from Photons}
The treatment of photons depends on an additional two transfer functions, $T_{\gamma\gamma}$ and $T_{c\gamma}$, defined in analogy to $T_{\gamma e}$ and $T_{ce}$, so that we have
\begin{equation} \label{eq:photonTF}
\begin{gathered}
    \frac{dN_\gamma^\text{out}}{dE} = T_{\gamma \gamma}(\delta, x_\mathrm{HI} | z,\Delta z) \frac{dN_e^\text{in}}{dE} \\
    \begin{bmatrix} \Delta T_k \\ \Delta x_e \\ J_\alpha \end{bmatrix} = D_{c\gamma}(\delta, x_\mathrm{HI} | z,\Delta z) \frac{dN_e^\text{in}}{dE},
\end{gathered}
\end{equation}
modeling the outgoing spectrum of photons and the contributions to the kinetic temperature, ionization fraction, and Ly$\alpha$ intensity from photon-scattering and associated processes between times $z$ and $z + \Delta z$. As in the case of electron energy deposition, these contributions are added to the time-evolution of the kinetic temperature, ionization fraction, and spin temperature performed by \cmfast in the \dmcm evaluation process.

Unlike electrons, the photons are not approximated as instantaneously depositing their energy, as their characteristic energy deposition length, though a function of photon energy, may be long on cosmological scales. In order to track the injection and deposition of energy by photons, \dmcm caches the local outgoing photon spectrum (summing spectra generated from both electron and photon scattering) at each time in the evolution for photon energies between $0.1-10\,\mathrm{keV}$. 

At each timestep in the evolution, the input spectrum is comprised of the local instantaneous injection spectrum and the emission history integrated along the past lightcone. Photons at energies above $10\,\mathrm{keV}$ have propagation lengths that are long, and so we can approximate photons at these energies produced in outgoing photon spectra as contributing to a spatially homogeneous bath that is evolved in time during the \dmcm evaluation. This homogeneous bath is also included at each timestep in the local instantaneous injection spectrum.

\subsubsection{Summary}
While this modeling procedure, which we have attempted only to survey as relevant, is somewhat involved, it is highly flexible. Given a prescription for the spatial dependence of the energy injection rate, $dE/dVdt$, and the spectrum of injected photons and electrons $dN/dE$, \dmcm can be directly applied to the imprints of that exotic energy injection of the 21-cm signal. In an expanded code release that builds upon that of \cite{Sun:2023acy}, we demonstrate how this may be applied to the physics scenarios considered in this work. For more details of the implementation and validation of \dmcm, see \cite{Sun:2023acy}.

\subsection{Projected Sensitivity Formalism}

In this section, we detail the procedure by which we calculate projected limits on exotic energy injection scenarios. 

\subsubsection{Background Model}
\label{sec:BkgModel}
To model the standard astrophysical and cosmological processes that determine the 21-cm signal in absence of exotic energy injection, we adopt the ``best-guess" scenario for 21-cm power spectrum modeling with \cmfast developed in \cite{Munoz:2021psm} and adopted in \cite{Sun:2023acy}. In this scenario, the reionization process is jointly driven by PopIII stars residing in the first-forming molecular cooling galaxies and later by PopII stars that form in atomic cooling galaxies. 

\begin{table}[!t]{
    \ra{1.3}
    \begin{center}
    \tabcolsep=0.08cm
    \begin{tabular}{c*{4}{C{0.07\textwidth}}}
    \hlinewd{1pt} 
    \textbf{PopII} parameters & $f_{\star,10}^\text{II}$ & $\alpha_\star^\text{II}$ & $f_\mathrm{esc,10}^\text{II}$ &  $L_X^\text{II}$ \\
    Fiducial value & -1.25 & 0.5 & -1.35 & 40.5 \\ \hlinewd{0.5pt}
    \textbf{PopIII} parameters & $f_{\star,7}^\text{III}$ & $\alpha_\star^\text{III}$ & $f_\mathrm{esc,7}^\text{III}$ &  $L_X^\text{III}$ \\ 
    Fiducial value & -2.5 & 0.0 &-1.35 & 40.5 \\ \hlinewd{0.5pt}
    \textbf{Shared} parameters & $t_\star$ & $\alpha_\mathrm{esc}$ & $E_0$ [eV] &  $A_\mathrm{LW}$ \\
    Fiducial value & 0.5 & -0.3 & 500 & 2.0 \\ \hlinewd{1.0pt}
    \end{tabular}\end{center}}
\caption{Summary of astrophysical parameters for reionization modeling used in this work. In our Fisher information treatment using \texttt{21cmfish}, each parameter is independently varied. For details, see Sec.~\ref{sec:BkgModel} and Refs.~\cite{Qin:2020xyh, Munoz:2021psm}. 
\label{tab:FisherNuisanceParameters}}
\end{table}

In \cmfast, the ionizing and heating ionizing and heating efficiencies of PopII and PopIII stars are determined by population-specific parameters $\{f_{\star,10}^\text{II}, \alpha_{\star}^\text{II}, f_\mathrm{esc, 10}^\text{II}, L_X^\text{II}\}$ and $\{f_{\star,7}^\text{III}, \alpha_{\star}^\text{III}, f_\mathrm{esc, 7}^\text{III}, L_X^\text{III},  A_\mathrm{LW}\}$, respectively, and the shared parameters $\{t_{\star}, \alpha_\mathrm{esc}, E_0\}$. The Lyman-Werner feedback on MCGs, affecting the PopIII star formation, is additionally described by the parameter $A_\mathrm{LW}$~\cite{Machacek:2000us}. We provide a table of our fiducial parameters in Tab.~\ref{tab:FisherNuisanceParameters}, and refer readers to \cite{Qin:2020xyh, Munoz:2021psm} for complete definitions. Additionally, we treat the impact of the DM-baryon relative velocity on the matter power spectrum and star formation efficiency in MCGs using the default implementation in \cmfast \cite{Munoz:2021psm}, taking parameters $A_{v_\text{cb}}=1$ and $\beta_{v_\text{cb}}=1.8$. We refer readers to \cite{Munoz:2021psm} for complete definitions and related discussions.

\subsubsection{Fisher Information Projections}
\label{sec:projection_method}
Our projected sensitivities are developed using the \texttt{21cmfish}\cite{Mason:2022obt} Fisher forecasting tool. Procedurally, for each astrophysical nuisance parameter $\theta_i$ summarized in Tab.~\ref{tab:FisherNuisanceParameters}, we perform a \dmcm evaluation without exotic processes at the fiducial value of that parameter, $\theta_i^\mathrm{fid}$, and at displaced values $\theta_i^\mathrm{fid} \pm \Delta \theta_i$, holding all other nuisance parameters fixed to their fiducial value, to develop predicted 21-cm power spectra. Similarly, for a parameter controlling the rate of exotic energy injection, which we denote $A$, we evaluate the power spectra at $A = 0$, $\Delta A$, and $2 \Delta A$.\footnote{In the case of \cite{Sun:2023acy}, this rate parameter was the dark matter decay rate, though in this work we consider alternate scenarios which require more general definition.} The size of the cosmological volume we simulate, resolution, time-stepping, and values of the cosmological parameters are all identical to those used in \cite{Sun:2023acy}. 

We then use \texttt{21cmfish} to evaluate the Fisher information matrix by taking the power spectrum generated under $A = 0$ with all nuisance parameters at their fiducial values to be the fiducial power spectrum and using the systematic variations of each parameter to calculate derivatives of the power spectrum and then of the likelihood with respect to each parameter of interest.\footnote{We have evaluated the energy injection rate at only nonnegative values of the parameter because the notion of negative energy injection is ambiguous. We then modify \texttt{21cmfish} to perform second-order accurate forward finite differences for the first and second derivatives with respect to $A$.} Our projections make use of the projected power spectrum at wavenumbers between $0.1\,\mathrm{Mpc}^{-1}$ and $1.0 \, \mathrm{Mpc}^{-1}$ as is expected to be measured by HERA, using 331 antennae for a total exposure of 1080 hours at $8\,\mathrm{MHz}$ radio frequency bandwidths between $50$ and $250\,\mathrm{MHz}$. We make use of the moderate foreground model developed in \texttt{21cmSense} \cite{pober201621cmsense}, and assume a HERA system temperature 
\begin{equation}
T_\mathrm{sys} = 100 \,\mathrm{K} + 120 \,\mathrm{K} \times \left(\frac{\nu}{150 \,\mathrm{MHz}} \right)^{-2.55},
\end{equation}
where $\nu$ is the observation frequency. We also include Poisson uncertainty and a 20\% modeling systematic uncertainty in our error budget following Ref.~\cite{Park:2018ljd}. These choices represent the astrophysics and uncertainty modeling used in the Fisher forecast of Ref.~\cite{Mason:2022obt} to reproduce the Bayesian analysis of Ref.~\cite{Munoz:2021psm} and then applied to dark matter decay scenarios in \cite{Sun:2023acy}. We report our projected 95$^\mathrm{th}$ percentile upper limit on the injection rate parameter as $1.65 \sqrt{[F^{-1}]_{AA}}$ where
$F$ is the fisher information matrix computed by \texttt{21cmfish}.

\section{Primordial Black Holes: Hawking Radiation}
\label{sec:pbh_evaporation}

A minimal scenario by which dark matter may drive exotic energy injection into the 21-cm signal is through the Hawking radiation of primordial black holes. Famously, black holes radiate, with a characteristic temperature inversely related to their mass, and so while stellar mass black holes have a very low temperature and associated emission rate, very light black holes with initial masses $M_\mathrm{PBH} \lesssim 10^{18}\,\mathrm{g}$ can radiate so efficiently that they would evaporate into energetic Standard Model particles on cosmological timescales, thereby leaving imprints on the thermal and ionization state of the early universe. While PBHs in this mass range are strongly excluded as comprising the whole of the DM, they are still viable as a small subcomponent of the DM. If such light black holes are indeed present in our universe, the absence of plausible late-universe production mechanisms suggests they would be primordial in nature ---formed in the very early universe as a result of either new high-scale physics or inflationary dynamics.

Very light PBHs, with initial masses $M_\mathrm{PBH} \lesssim 10^{15}\,\mathrm{g}$ represent an even more extreme scenario, with the PBHs having fully evaporated by the present time. As a result, these scenarios are sensitively probed by searches for exotic energy injection such as 21-cm cosmology, though they may also be probed by CMB distortions or even through their possible remnants; see \cite{Dvali:2018xpy, Dvali:2018ytn,Dvali:2020wft, Alexandre:2024nuo, Dvali:2024hsb, Thoss:2024hsr, Montefalcone:2025akm}. In particular, although these light PBHs may evaporate rapidly, the high energy photons they generate deposit their energy relatively slowly, leading to later-time heating and ionization of the IGM at redshifts probed by 21-cm observations. 

In this section, we present the details of our energy injection modeling (in Sec.~\ref{sec:EvapImplementation}) and develop forecast constraints on monochromatic PBH scenarios (in Sec.~\ref{sec:EvapResults}). Note that in this work, we assume, when allowed by the emission rate, complete evaporation of light PBHs, though effects like the memory burden may cut off the evaporation process, leading to somewhat weakened projected sensitivities \cite{Dondarini:2025ktz}, though see \cite{Montefalcone:2025akm}.

\subsection{Implementation in \dmcm}
\label{sec:EvapImplementation}

In the \dmcm modeling framework, we write the energy injection by PBH evaporation as in Eq.~\ref{eq:EnergyInjectionRate} by 
\begin{equation}
\begin{split}
    \frac{dN_{\gamma/e}}{dE dV dt} = (1 + \delta) f_\mathrm{PBH} \rho_\mathrm{DM} \int dM &\frac{dN}{dM} \Gamma(M,z) \\
    \times & \frac{dN_{\gamma/e}}{dE}(M,z)
\end{split}
\end{equation}
where $dN/dM$ is the unit-normalized initial PBH mass function, $\Gamma$ is the injection rate of the hadronized photon/electron spectra $dN_{\gamma/e}/dE$. The fractional abundance of PBHs, $f_\mathrm{PBH}$, is defined relative to the DM abundance $\rho_\mathrm{DM}$.\footnote{The mass range of PBHs we consider in this work includes masses for which the PBH lifetime is less than the age of the universe. In these scenarios, the present-day fractional abundance of PBHs is then 0. Then the PBH fraction $f_\mathrm{PBH}$, which we consider in this work, is defined by the initial PBH energy density relative to the Planck18 DM density back-scaled to the time of formation.} For each initial mass $M$, we use \texttt{BlackHawk v2.3} \cite{Arbey:2019mbc, Arbey:2021mbl} to obtain the emission rate $\Gamma(z)$ and a primary spectrum $dN_i(z)/dE$ where the $i$ index runs over all Standard Model particles and also gravitons. We generate the primary spectra over 20000 log-uniform energy bins between $10^{-5}\,\mathrm{eV}$ and $100\,\mathrm{TeV}$; further increasing the energy range and resolution does not change the results presented here. There have also been studies that consider $O(\alpha)$ dissipative corrections to the Hawking radiation, leading to differing predictions for the soft photon (but not hard photon) spectrum which could be incorporated in more detailed future analyses \cite{Koivu:2024gjl, Silva:2022buk}.

\begin{figure}[h!]
    \centering
    \includegraphics[width=\linewidth]{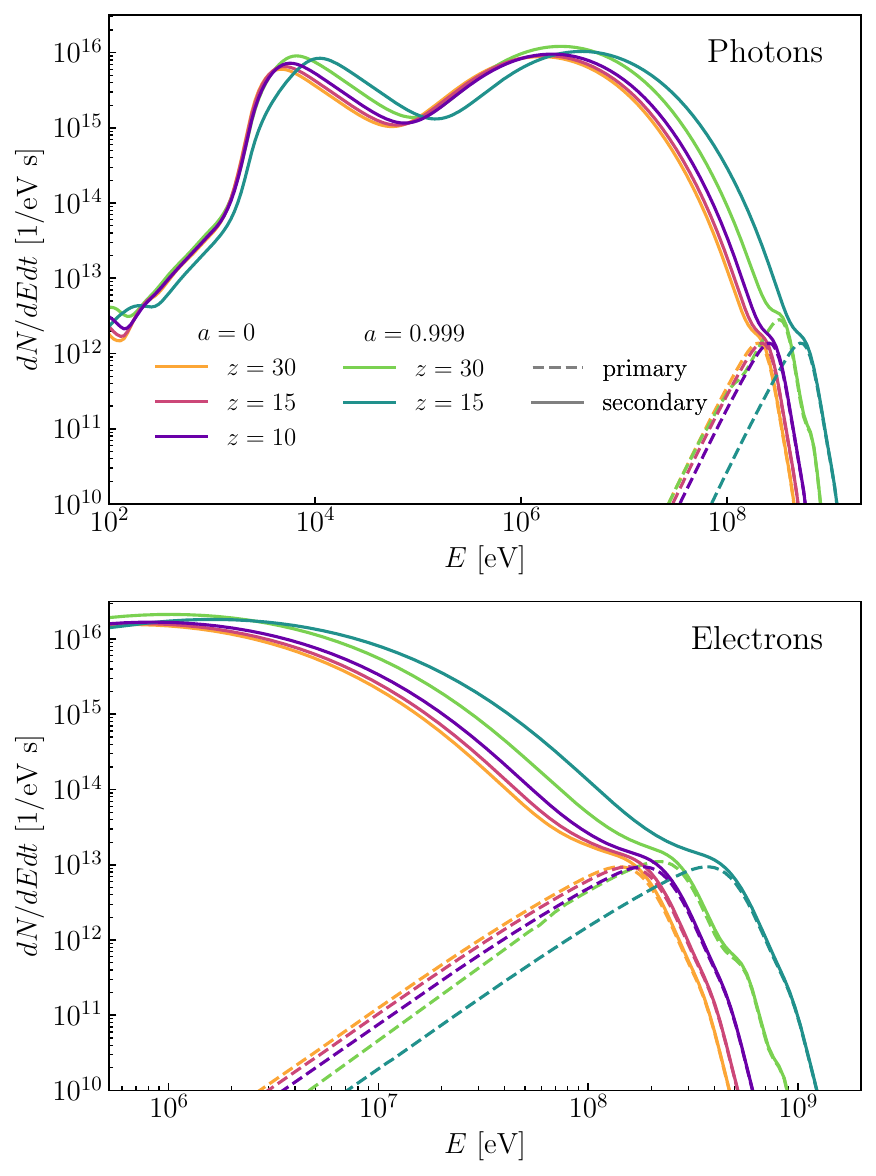}
    \caption{\textbf{Example Hawking radiation spectra of photons and electrons} of PBHs with mass $10^{14.5}$ g, in the zero spin ($a=0$) and the high spin ($a=0.999$) cases. The dashed lines show the primary radiation, while the solid lines show the final injected electron and photon spectra after hadronization. Spectra with different colors correspond to different redshifts, which change significantly as the zero (high) spin PBH is nearing evaporation at $z=6.75$ ($z=13.7$). Large mass PBHs realize less extreme time-variation in their spectra due to their slower evaporation rate.
    }
    \label{fig:pbh-hr-example-specs}
\end{figure}

For consistency with \dhis and \dmcm, we use the \texttt{PPPC4DMID} hadronization table supplied in \dhis to hadronize the primary particles, instead of the hadronization modules provided in \bh \cite{Cirelli:2010xx}. We also find that \texttt{PPPC4DMID} realizes better energy conservation in the sense that the total energy of the secondary states matches the total energy of the primary spectra than \bh. The \texttt{PPPC4DMID} tables relate the input energy $E_\mathrm{in}$ of a single primary spectrum particle to $dN(E_\mathrm{in})/dx$, the spectrum of secondary particles it produces, with $x = E_\mathrm{out}/E_\mathrm{in}$. As the \texttt{PPPC4DMID} tables are generated with \texttt{PYTHIA} only for primary particles with with energies $E_\mathrm{in}$ between $5\,\mathrm{GeV}$ and $100\,\mathrm{TeV}$, we extrapolate the tables by assuming $d{N}(E_\mathrm{in})/dx$ evaluted at primary spectrum energies below $5\,\mathrm{GeV}$ to $d{N}(E_\mathrm{in} = 5\,\mathrm{GeV})/dx$. Note that because our primary spectra may be time-dependent through the potentially appreciably changing PBH mass over cosmological history, we must generate our secondary spectra within each time step of the \dmcm evolution. In Fig.~\ref{fig:pbh-hr-example-specs}, we present the primary and secondary spectra of electrons as a function of redshift for PBHs of initial mass $10^{14.5}\,\mathrm{g}$, in the zero spin and the high spin case. It is illustrative of the overall time-dependence as they evaporate by $z \approx 6.75$ and $z \approx 13.7$ respectively.

This time-dependence represents a new challenge compared to the gradual and constant decay rate scenarios considered in \cite{Sun:2023acy}, as when the PBHs near evaporation, their decay rates accelerate to large values. If not treated carefully, our \dmcm evolution of Eq.~\ref{Eq:ODE} will become inaccurate when its timestep $\Delta t$ becomes large compared to the inverse decay rate. To address this issue, when we perform a timestep evolving the thermal and ionization state from $t$ to $t + \Delta t$, we inject not the instantaneous but the integrated secondary spectrum. Additionally, we have confirmed that the results presented in this work do not change when the timestep is further decreased in size, indicating good convergence.

\subsection{Results}
\label{sec:EvapResults}

\begin{figure}[h!]  
    \begin{center}
    \includegraphics[width=\linewidth]{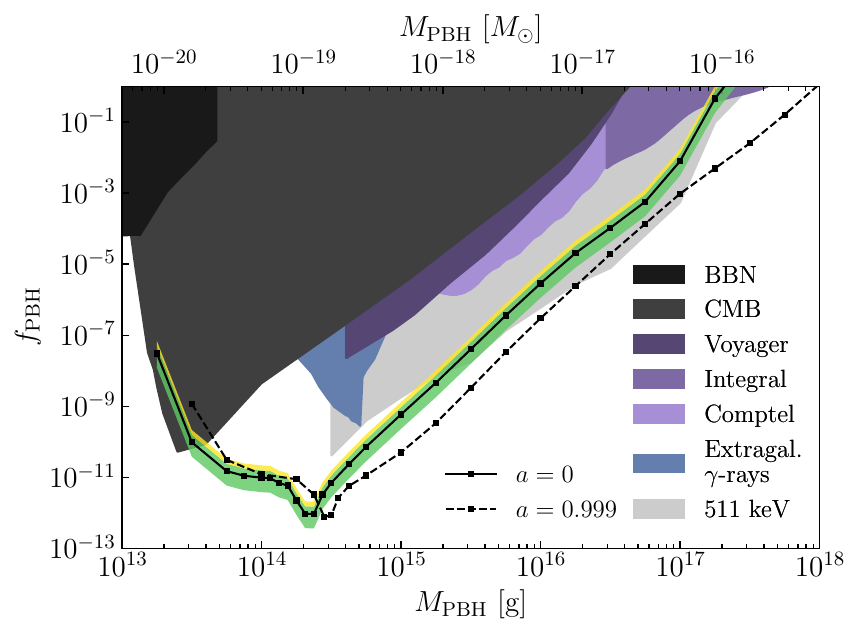}
    \caption{
    \textbf{Projected 95$^\text{th}$ percentile upper limits on evaporating PBHs.} 
    The solid line represents our projected limits for zero spin ($a=0$) PBHs, while the dashed line represents those for a high spin case ($a=0.999$). The green and yellow bands correspond to the 1 and 2 $\sigma$ uncertainty on the limit, respectively. A similar band for the high spin case is omitted for clarity. Bumps in sensitivity occur between $10^{14}$-$10^{14.5}$~g when the final evaporation time of the PBHs coincide with the start of reionization at $z\sim15$. Towards the small masses, the sensitivity quickly decreases as the PBHs fully evaporate earlier and earlier, producing only global heating and ionizing signatures. We also include limits from BBN \cite{Carr:2009jm}, CMB \cite{Clark:2016nst, Acharya:2020jbv}, Voyager \cite{Boudaud:2018hqb}, Integral \cite{Laha:2020ivk,Berteaud:2022tws}, Comptel \cite{Coogan:2020tuf}, extragalactic $\gamma$-rays \cite{Carr:2020gox}, and 511~keV line \cite{DelaTorreLuque:2024qms}, some of which are compiled on PBHBounds \cite{PBHbounds}.
    }
    \label{fig:pbh_evap_pro}
    \end{center}
\end{figure}

We develop projected sensitivities to $f_\mathrm{PBH}$ following the procedure defined in Sec.~\ref{sec:projection_method}. We assume a monochromatic PBH mass function, and consider initial masses between $10^{13}\,\mathrm{g}$ and $10^{18}\,\mathrm{g}$. We also consider two values of the PBH spin parameter, denoted $a$, and not to be confused with the scale factor. As a bracketing interval, we consider $a = 0$ and $a = 0.999$,  with $a = 1$ corresponding to maximal spin. Our projected sensitivities are summarized in Fig.~\ref{fig:pbh_evap_pro}.  

Maximal projected sensitivity is realized for $M_\mathrm{PBH} =10^{14}\sim10^{14.5}$~g such that  the final stages of the PBHs' evaporation coincide with reionization. Since the evaporation time is spin-dependent, the sensitivity peaks at $M_\mathrm{PBH}\approx 2.2\times10^{14}$~g for zero-spin PBHs, but $M_\mathrm{PBH}\approx 3.0\times10^{14}$ for the high-spin ($a = 0.999$) case. This feature is consistent with our understanding that deposition near $z=10\sim15$ maximally impacts the 21-cm brightness temperature. These redshifts are accessible to 21-cm observatories like HERA but are essentially uncontaminated by stellar energy injection. Proceeding to lower PBH masses, the complete evaporation occurs at even earlier times, leading to energy injection and deposition at epochs unprobed by 21-cm observatories. At higher masses, the slower evaporation rate leads to less total energy injection at redshifts accessed by 21-cm observatories and correspondingly weaker limits. If realized, these sensitivities would represent the strongest constraints on light PBH masses over a broad range of masses.

\section{Primordial Black Holes: Accretion}
\label{sec:pbh_accretion}
While very light PBHs can drive exotic energy injection in the early universe through Hawking radiation, more massive PBHs do not generate appreciable energy injection through this process due to their very low Hawking temperatures. On the other hand, more massive (\textit{i.e.}, stellar mass) black holes present in the early universe can efficiently accrete baryonic matter, with infalling matter being heated at the horizon to temperatures as large as $10^{11\sim12}$~K \cite{B52, XY12, PR13, YN14}. As these temperatures correspond to energy scales of up to hundreds of MeV, the accretion process causes \textit{X}-ray and $\gamma$-ray emission that contribute to early heating and ionization. The role of accretion-powered emission has been previously studied in terms of its imprints on the CMB \cite{AK17, PSCCK17, FLC24, AEGSSV24} and on the global 21-cm signal \cite{Yang:2021idt, Zhang:2025xjs}, and in this work, we extend its study to the context of the 21-cm power spectrum, also studied in \cite{Mena:2019nhm}. Also, though we consider only PBHs, a more detailed treatment of black hole accretion might be relevant even for 21-cm cosmology in standard scenarios, as astrophysical BHs realized as the endpoint of stellar evolution are expected to contribute to IGM heating at cosmic dawn.

As a whole, the process of exotic energy injection from baryonic accretion onto PBHs is a highly inhomogeneous process, making its study in \dmcm particularly appropriate. PBHs form in the presence of large local overdensities, resulting in an early onset of matter accretion and recombination  \cite{PSCCK17}. The accretion efficiency of those PBHs then depends heavily on their baryonic environment, particularly through its dependence on the gas streaming velocity and temperature in the host halo of the PBH during the late dark ages \cite{PR13I, PR13II, PR13}. Finally, the energy injection, which proceeds mostly through free-free emission, has support at photon energies ranging from very low energy photons to soft gamma-ray photons, which experience drastically different opacities and cooling lengths as they deposit their energy \cite{Sun:2023acy}. This in turn allows for both highly localized and non-local energy deposition from a single accreting PBH. 

In this section, we begin with a review of how the luminosity of an accreting black hole is evaluated in terms of its velocity relative to the baryonic matter and the gas density and sound speed in Sec.~\ref{sec:BHLuminosity}, followed by a discussion in Sec.~\ref{sec:BHEnvironment} of how each of these quantities can be calculated as relevant in a environmentally dependent and spatially varying manner using input \cmfast simulation states as evolved in \dmcm. We then examine the impact of the modeling choices made in treating energy injection via black hole accretion before presenting constraints on the PBH as a function of mass in Sec.~\ref{sec:BHModelVariationsResults}.

\subsection{Modeling the Black Hole Luminosity}
\label{sec:BHLuminosity}
Here, we define the procedure by which we model accreting black holes. We discuss models for the mass accretion rate in Sec.~\ref{sec:accretion-modeling}, followed by a prescription for relating the mass accretion rate to the accretion-powered luminosity and spectrum in Sec.~\ref{sec:luminosity_eff_spec}.

\subsubsection{Black Hole Accretion Models}
\label{sec:accretion-modeling}

The two benchmark models for the accretion rate of black holes we consider in this work are the classical Bondi-Hoyle-Lyttleton (BHL) model \cite{B52, HL39} and the more recent simulation-informed Park-Riccoti (PR) model \cite{PR13I, PR13II, PR13, SR20}. We briefly survey and comment on each here. We note that, in general, there exist considerable theoretical uncertainties in understanding the PBH accretion rates with appreciable disagreement in the literature, see, \textit{e.g.},  \cite{B52, AK17, PSCCK17, FLC24, AEGSSV24}. We do not argue for either model in particular, and we instead seek to benchmark the projected sensitivities associated with either BHL- or PR-like accretion. We do, however, note that the PR model predicts generally lower rates, leading to more conservative sensitivity estimates from the 21-cm power spectrum, analogous to results in the context of the CMB as studied in \cite{AEGSSV24}.

The classical Bondi-Hoyle-Lyttleton (BHL) model \cite{B52, HL39} interpolates between the accretion scenario of a moving BH in a pressure-less gas, and that of a stationary BH accreting gas with pressure taken into account. The accretion rate onto a black hole is given by
\begin{equation} \label{eq:BHL-acc-rate}
\dot{M}_\text{BHL}=4\pi\lambda\rho_{\infty}\frac{G^2 M^2}{(v_\infty^2+c_{\text{s},\infty}^2)^{3/2}},
\end{equation}
where $\rho_{\infty}$, $c_{\text{s},\infty}$ are respectively the baryon density and sound speed away from the BH, and $v_\infty$ is the relative velocity of the BH to the gas far away. Following \cite{AEGSSV24}, we have also included a phenomenological correction factor of $\lambda$, which seeks to account for sub-BHL accretion rates inferred at a variety of observational targets. In this work, like in \cite{AEGSSV24}, we assume $\lambda = 0.01$ as our benchmark value when considering BHL accretion, and we denote the total modeling procedure (which also includes emission and environmental effects) when performed with the BHL accretion model by \BHL.

A more simulation-informed modeling prescription for black hole accretion rate is provided by the PR model. Simulations of intermediate mass black holes ($M\sim10^{2\sim5}\Msun$) and at large gas densities $n_\text{H}\sim10^{2 \sim 5}$/cm$^3$ find that the formation of an ionization region / Str\"omgren sphere around the black hole causes the accretion rate to significantly deviate from BHL model prediction, particularly for black holes which moving with a velocity that is less than the sound speed. The PR model finds that the accretion rate is better described instead by 
\begin{equation} \label{eq:PR-acc-rate}
    \dot{M}_\text{PR}=4\pi\rho_\text{in}\frac{G^2 M^2}{(v_\text{in}^2+c_\text{s,in}^2)^{3/2}}.
\end{equation}
where $\rho_\text{in}$ and $c_\text{s,in}$ are the baryon density and sound speed inside the ionized region, and $v_\text{in}$ is the BH's relative velocity to the baryons in the ionized region. 

In the PR model, the values of $\rho_\text{in}$ and $v_\text{in}$ are determined from the conservation of mass and momentum across the I-front, and $c_\text{s,in}$ is a model parameter that depends on the detailed feedback in the ionized region. The conservation equations admit two solutions for $\rho_\text{in}$ and $v_\text{in}$, corresponding an densified (\textit{D}-type) I-front, and a rarefied (\textit{R}-type) I-front. The PR model finds that a \textit{D}-type I-front forms when the BH is moving slower than $v_D\approx c_{\text{s},\infty}^2/2c_\text{s,in}$, which is always subsonic, and that a \textit{R}-type I-front forms when the BH is moving faster than $2c_\text{s,in}$. In the intermediate velocity range, it is observed from simulation that almost always, $v_\text{in}\approx c_\text{s,in}$.  For details on how how $\rho_\mathrm{in}$ and $v_\mathrm{in}$ are calculated, see App.~\ref{app:pr_model}. 

Since the PR model is better calibrated against simulations performed for black hole masses in a range that overlaps with the range of masses of interest for this work, we adopt it as our fiducial accretion model, and denote our fiducial energy injection (including accretion and emission) model as \PR. However, some caution is required since the simulations that inform the PR model were performed at larger gas densities than realized during the cosmic dark ages and the epoch of reionization.

Following Ref.~\cite{AEGSSV24}'s implementation of the model, we also choose $c_\text{s,in}=23$~km/s as the fiducial interior sound speed, which corresponds to a temperature of $4\times10^4$~K in the ionized region \cite{SR20}. The interior sound speed $c_\text{s,in}$ is a parameter in the PR model that is in reality determined by the detailed heating and cooling mechanisms of the ionized region surrounding the BH, dependent on the gas density and BH mass \cite{SR20}. While we do not make a model-informed calculation of this parameter, we choose two systematic variations of its value, taking $c_\text{s,in}=14$~km/s and $c_\text{s,in}=29$~km/s to show the impact on the PBHs' luminosity. We refer to these variations by \PRcm and \PRcp, respectively. These sound speeds correspond to interior temperatures of $1.5\times10^4$~K and $6\times10^4$~K, which approximately bracket the observed temperatures in simulations in \cite{SR20}. For more detailed descriptions of the PR model, we refer the readers to the Ref.~\cite{AEGSSV24} and the original references Ref.~\cite{PR13, SR20}.

\subsubsection{Luminosity Efficiency and Spectrum}
\label{sec:luminosity_eff_spec}
While the black hole accretion rate can be directly estimated, under the assumption of either the PR or BHL models, from the gas velocity and density, characterizing the emission luminosity associated with the gas accretion requires careful modeling of small-scale physics in the vicinity of the BH horizon \cite{YN14, AK17}. While emission associated with spherically symmetric accretion was studied in Ref.~\cite{AK17}, Ref.~\cite{PSCCK17} instead claims that baryons are likely to form accretion disks around PBHs, thereby spoiling the spherical symmetry. The key argument of Ref.~\cite{PSCCK17} is that the accretion disk will form when the characteristic angular momentum of accreted gas is greater than the angular momentum associated with the innermost stable circular orbit (ISCO). Parametrically, the condition for disk formation is then 
\begin{equation} \label{eq:disk-criterion}
    l\approx\left(\frac{\delta\rho}{\rho}+\frac{\delta v}{v }\right) v~r_\text{B}>l_\text{ISCO},
\end{equation}
where $\delta\rho/\rho$ and $\delta v/v $ are the relative inhomogeneities of gas density and velocity on the scale of the accretion problem, respectively, and $r_B$ is the typical impact parameter from which the infalling matter originates, corresponding to the Bondi radius in the BHL accretion model. 

The authors of Ref.~\cite{PSCCK17} argued that, for unbound PBHs, Eq.~\ref{eq:disk-criterion} is satisfied with density perturbation of $\delta\rho/\rho\gg10^{-4}$ which is satisfied due to the presence of PBHs, while for halo-bound PBHs, is it satisfied for the velocity variation $\delta v\gg1.5$~m/s after recombination, which in turn can be easily satisfied in a halo environment with km/s scale velocity dispersion. For reference, the velocity dispersion at scale radius for a 100 $\Msun$ halo at $z=50$ is $\sim300$~m/s, see App.~\ref{app:vDisp}. Thus, we assume that accretion disks always form around both halo-bound and unbound PBHs. We then treat all accretion processes as associated with the formation of accretion disks in both the \textit{thin disk} and \textit{thick disk} regimes by relating the accretion luminosity to the accretion rate as
\begin{equation}
    L = \epsilon \dot{M} c^2.
\end{equation}
using an efficiency parameter $\epsilon$.

BHs that accrete at a rate close to the Eddington accretion rate, defined by
\begin{equation}
    \dot{M}_\mathrm{Edd} =10\cdot\frac{4 \pi G M m_p}{\sigma_T c}
\end{equation}
where $M$ is the mass of the accreting black hole, $m_p$ is the proton mass, and $\sigma_T$ is the Thompson cross-section, are known to be well-described by the \textit{thin} disk model with $\epsilon \approx 0.1$ \cite{XY12, YN14}. For more slowly accreting black holes, the accretion disk that forms is geometrically thicker but less dense, leading to less efficient radiation of accreted mass-energy through processes involving electrons. In this scenario, the accretion disk dynamics can be described by a hot \textit{Advection-Dominated Accretion Flow} (ADAF), in which energy released due to gas turbulence is deposited mainly in the ionic component of the hot gas, with only a fraction $\delta_e$ going to electrons and can be radiated efficiently, resulting in a lowered emission efficiency.

\begin{figure}[!t]
    \centering
    \includegraphics[width=\linewidth]{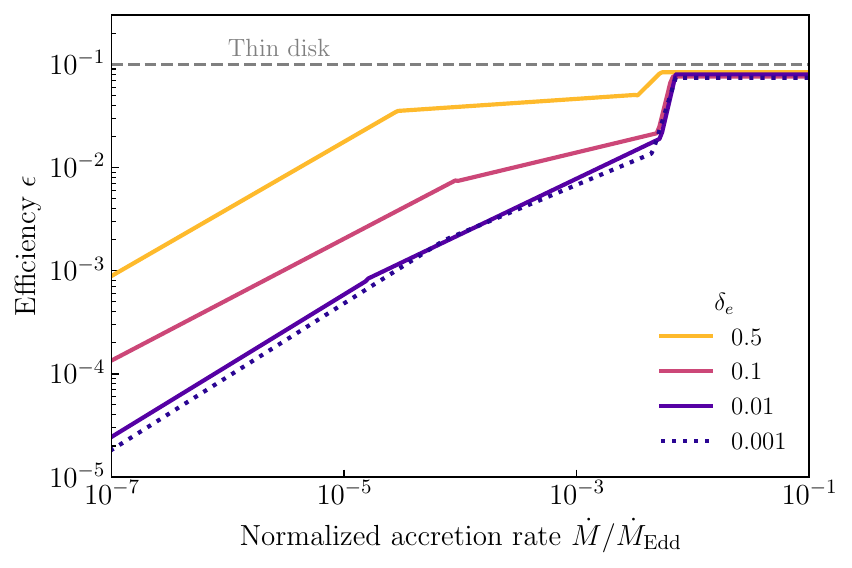}
    \caption{\textbf{Fitted ADAF emission efficiency as a function of $\delta_e$.} Adapted from Tab.~I of Ref.~\cite{XY12}, which numerically solves the ADAF equations and fitted piecewise polynomial functions to the solutions. The solid lines show efficiency for electron heating fractions $\delta_e=0.5,0.1,0.01$ respectively, corresponding to our \PRdp, \PR, \PRdm variations. The case for $\delta_e=0.01$ is similar to the dotted line showing $\delta_e=0.001$, as it is approaching a lower limit of emission efficiency where direct viscous heating controlled by $\delta_e$ is no longer dominant \cite{XY12}. See text for details.}
    \label{fig:e-Mratio}
\end{figure}

The relationship between the electron heating fraction $\delta_e$\footnote{We have renamed the electron viscous heating fraction $\delta_e$, instead of $\delta$, which is common in literature, to avoid confusion with the matter overdensity $\delta$.} and the emission efficiency was studied in Ref.~\cite{XY12} by numerically solving the 1D ADAF equations. The results are then fitted to a piecewise power-law parametrization of the form 
\begin{equation}
    \epsilon=\epsilon_0\left(\frac{\dot M}{0.01 \dot M_\text{Edd}}\right)^a,
\end{equation}
with $a$ and $\epsilon_0$ provided in Tab.~1 of \cite{XY12} as functions of $\delta_e$. While debate remains about the correct value for $\delta_e$, a value of $0.1\sim0.5$ is generally accepted \cite{XY12}. We conservatively choose $\delta_e=0.1$ as our fiducial choice, and consider choices of $\delta_e=0.5$ and $\delta_e = 0.01$ as model variations \PRdp, \PRdm respectively. These efficiencies are illustrated in Fig.~\ref{fig:e-Mratio} as functions of the normalized accretion rate $\dot{M}/\dot{M}_\mathrm{Edd}$. Note that \cite{XY12} additionally studied the efficiency for $\delta_e=0.001$, which is very similar to the $\delta_e=0.01$ case, suggesting that emission persists despite the inefficient viscous heating of electrons, making the $\delta_e=0.01$ case a lower limit on the ADAF emission efficiency. Additionally, Fig.~\ref{fig:e-Mratio} shows that this ADAF model very nearly realizes the thin disk limit of accretion efficiency at large accretion rates. Moreover, we find that in the contexts studied in this work we very rarely encounter scenarios in which accretion becomes comparable to the Eddington accretion rate. As a result, we consistently adopt the ADAF emission efficiency model. 

As the emission in the ADAF model arises mostly from electron free-free scattering, the spectrum is expected to be relatively flat, with a high-energy cutoff at temperatures close to the PBH surface $T_s\sim m_e$, and a lower cutoff $E_\text{min}$ determined by the opacity of the accretion disk itself. Similar to Ref.~\cite{PSCCK17}, we will use the numerical simulations of ADAF models shown in Fig.~1 of Ref.~\cite{YN14}, parametrized in Ref.~\cite{PSCCK17} as
\begin{equation}
    \frac{dL}{dE}\propto\Theta(E-E_\text{min})E^{-a}\exp(-E/T_s),
\end{equation}
where $E_\text{min}$ is given by $E_\text{min}=(\Mpbh/10M_\odot)^{-1/2}$~eV. Spectra in simulations exhibit scaling ranging in $a\in[0.7, 1.3]$. We will take $a=1$ and $T_s=200$~keV for concreteness, as in Ref.~\cite{PSCCK17}.

\subsubsection{Summary}

Altogether, we can summarize our PBH emission modeling as follows. Given the choice of accretion model, the bolometric luminosity of accretion-powered emission depends on the $\Mpbh$ and the accretion rate $\Mdotpbh$, which depend on the ambient gas density $\rho_\infty$, streaming speed $v_\infty$, and the sound speed $c_{s,\infty}$. The spectral shape of the emission is fixed except for the $\Mpbh$ dependent lower cutoff. In all, the luminosity of a PBH given a fixed model can be written as
\begin{equation} \label{eq:dLdE}
    \frac{dL}{dE}=\frac{dL}{dE}(\Mpbh, \rho, v,  c_s).
\end{equation}
where $\rho$, $v$, and $c_s$ are the ambient baryon density, velocity, and sound speed. We now proceed to describe how each of these ingredients can be calculated in an environmentally dependent way using a combination of the \cmfast pipeline and subgrid modeling.

\subsection{Environmental Inputs for BH Accretion}
\label{sec:BHEnvironment}

In order to calculate the total energy injection, we must combine the total accretion luminosity and luminosity spectrum, detailed in Sec.~\ref{sec:BHModelVariationsResults}, with environmental modeling that specifies the number of accreting primordial black holes and the inputs for the accretion luminosity modeling. These inputs, which are environmentally dependent and therefore generally spatially varying, can be determined by considering PBH populations in one of two possible environments:
\begin{itemize}
    \item Unbound PBHs which are not housed within any gravitationally collapsed halos
    \item Halo-bound PBHs which reside within larger gravitationally collapsed halos
\end{itemize}
We also consider the possibility that ultracompact minihalos (UCMHs) form in the neighborhood of PBHs (either unbound or halo-bound). However, we consider this possibility only as a systematic variation on our fiducial modeling, which neglects it, because the prospects for UCMH formation in the neighborhood of a PBH, which depends on self-similar radial infall absent any angular momentum or environmental perturbing effects, is more speculative in nature, and because the baryonic abundance in those UCMHs, which are conventionally studied in the context of collisionless dark matter, is poorly understood \cite{Delos:2017thv, Delos:2018ueo}.

\subsubsection{Unbound PBHs}
\label{sec:unbound_pbh}

We begin by considering accretion-powered emission onto PBHs which do not yet reside in halos, whether because the accretion occurs at early times before halos form or because the PBH resides in a local underdensity that experiences a delayed onset of structure formation. Indeed, it is these unbound PBHs that are primarily responsible for early-time energy injection ($z \gtrsim 100$, well before appreciable structure formation has occurred), resulting in CMB sensitivities to accreting PBHs \cite{AK17, PSCCK17, FLC24, AEGSSV24}. 

Since these PBHs are unbound to halos, we take the local baryon density to be that of the large-scale cosmological average  \cite{PSCCK17}, and we take the relative velocity between the PBHs and baryons to be determined by the DM-baryon streaming velocity, which arises as baryons decouple from acoustic oscillations \cite{Tseliakhovich:2010bj}. Numerical solutions to the Boltzmann equations suggest this streaming velocity follows a Maxwell-Boltzmann distribution with an RMS velocity of 
\begin{equation} \label{eq:streaming-v}
    \sqrt{\langle v_\text{rel}^2\rangle}=\text{min}\left(1,\frac{1+z}{1000}\right)\cdot 30~\text{km/s},
\end{equation}
though any enhanced small-scale structure and associated velocity fluctuations could modify this expectation \cite{Dvorkin:2013cea}. Since neglecting any small-scale enhancements represents a conservative assumption for the accretion luminosity, we make it our fiducial modeling choice, though we revisit this when considering the role of UCMHs around PBHs in Sec.~\ref{sec:UCMHs}.

The DM-baryon streaming velocity $v_{\text{cb}}$ is initialized in \cmfast at the redshift of decoupling $z_\mathrm{dec}$, by generating each of its three coordinate components (keeping correlations with the density field) and squaring them. The streaming velocity is assumed to redshift as $(1+z)$ thereafter \cite{Munoz:2019rhi, Munoz:2021psm}. We will take the redshifted value to be the input gas-BH relative velocity in our accretion model
\begin{equation}
    v_{\infty,i}=\frac{1+z}{1+z_\text{dec}}\,v_{\text{cb},i}(z_\mathrm{dec}),
\end{equation}
where $i$ represents the $i$-th cell and $z_\mathrm{dec} \approx 1060$.  Our fiducial simulation cell has a side length of 2 conformal Mpc, and it is assumed that the unbound PBHs stay within a cell, which is a good approximation given that the overdensity, temperature, and ionization fraction are mostly smooth on the scale of the cell size. Problems may arise when halos that are more massive than a cell form; however, as we will discuss in Sec.~\ref{sec:HaloBoundPBHs}, we do not need to worry about such large halos until nearing the end of the simulation at $z\sim5$, at which any energy injection stops contributing significantly to detectable brightness temperature power spectrum. For the baryon density experienced by unbound PBHs, we discount the fraction of matter collapsed into halos, and are left with
\begin{equation}
    \rho_{b,i}=(1-f_{\text{coll}}(\delta_i))(1+\delta_i)\overline{\rho_b}
\end{equation}
where $\delta_i$ is the overdensity of the cell, and $f_\text{coll}$ the collapsed fraction, computed with the Sheth-Tormen halo mass function consistent with \cmfast \cite{2011MNRAS.411..955M}. The ambient baryon sound speed at each cell is given by the local matter temperature $T_{k,i}$ and ionization fraction $x_{e,i}$ within each cell
\begin{equation}
    c_{\infty,i}=\sqrt{\frac{5}{3}(1 + x_{e,i})\,T_{k,i}}.
\end{equation}
Given these environmental quantities, we can calculate the PBH emission luminosity and spectrum in each cell of the simulation from Eq.~\ref{eq:dLdE}. In \dmcm, we precompute the injected power as a tabulated function of overdensity $\delta$, streaming velocity $v_\infty$, and ambient sound speed $c_\infty$, which we then linearly interpolate in the simulation.

To set the scale of the importance of unbound halo accretion, we consider the accretion-powered luminosity of an unbound PBH at $z= 15$, around the time that we are most sensitive to exotic energy injection through the 21-cm power spectrum. At this time, the collapse fraction in halos of mass greater than 1 $\Msun$ is $f_\text{coll}\approx0.17$, so there are roughly comparable numbers of PBHs that are halo-bound as compared to unbound.\footnote{Halos which are at least 30 times more massive than the PBH mass are assumed to host PBHs, which we discuss at greater length in Sec.~\ref{sec:HaloBoundPBHs}} At $z=15$, the diffuse gas density around our PBH is around $8.4\times10^{-4}$/cm$^3$ with temperature expected to be cooled to $5.5$~K without star-formation heating, which corresponds to a low sound speed of $0.27$~km/s. The streaming velocity has also redshifted to a low value, typically $0.48$~km/s. We assume our fiducial PR model with interior sound speed at $c_s=23$~km/s. Our PBH will sit in the intermediate velocity regime (see App. A of Ref.~\cite{AEGSSV24}), and the immediate surround gas density very suppressed at $2.4\times10^{-7}$/cm$^3$. Taking a 100 $\Msun$ PBHs as an example, the accretion rate is around $5.1\times10^{-19}\Msun/$yr, with the total emission luminosity around $1.8\times10^{21}$~erg/s, corresponding to an efficiency of $\epsilon=6.3\times10^{-8}$. Overall, this is a very low luminosity. As we will see in Sec~\ref{sec:HaloBoundPBHs}, the PBHs that are bound within halos make a much larger contribution to the accretion-powered energy injection at this time. 

\subsubsection{Halo-bound PBHs}
\label{sec:HaloBoundPBHs}
At redshifts $z \lesssim 50$, a non-negligible fraction of the DM and baryons begin to collapse into halos which realize much denser gas densities than the large-scale cosmological average, in turn boosting the accretion rate of PBHs which reside within those halos as compared to unbound PBHs. As we will demonstrate, the precise value of the enhancement relative to the unbound PBH accretion rate is highly dependent on the choice of either PR or BHL accretion models, but for all modeling choices studied in this work, energy injection from halo-bound PBH accretion dominates over the unbound PBH contribution for all redshifts relevant for 21-cm power spectrum sensitivities.

To calculate the bound PBH accretion luminosity, we begin by characterizing the baryonic environment generated by early forming halos. We assume that both the DM and the baryons are distributed following an NFW profile
\begin{equation}
    \rho_\mathrm{NFW}(r)\propto\frac{1}{r/r_s(1+r/r_s)^2},
\end{equation}
where $r_s$ is the NFW scale radius. The halo mass is defined by integrating the total density out to $r_\Delta$, the radius at which the average density inside is $\Delta$ times the critical density. We take $\Delta=200$, and use the analytical halo mass-concentration relation calibrated against numerical simulations studied in Ref.~\cite{Ludlow:2016ifl} and compiled in \texttt{halomod v2.2} \cite{Murray:2013qza, Murray:2020dcd}. Given the concentration, the scale radius and density normalization can be determined. We assume that across the halo, the baryon-DM ratio follows that of the cosmological average. This is a simple modeling assumption that could be made more robust with sharper predictions for the baryonic matter distribution in early forming halos, including those that host star formation.  

\begin{figure}[!ht]
    \centering
    \includegraphics[width=\linewidth]{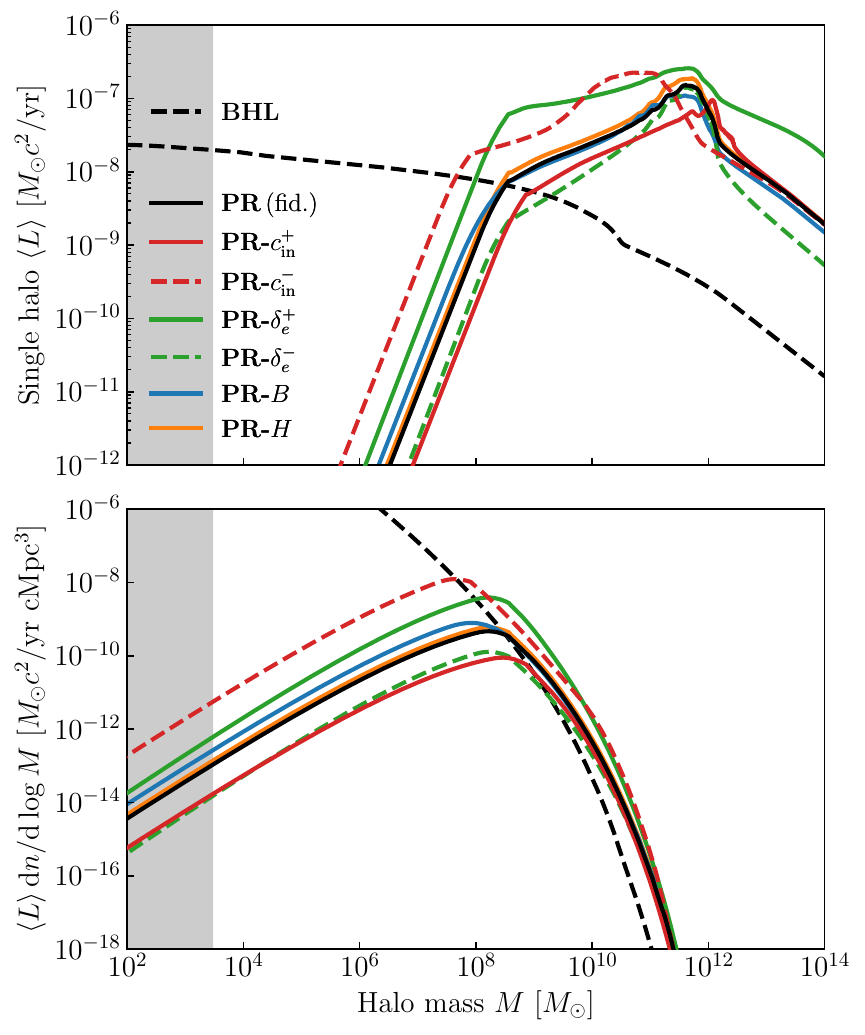}
    \caption{\textbf{Halo-integrated PBH luminosity as a function of halo mass} at redshift $z=15$, for $M_\mathrm{PBH}=100\Msun$, $f_\mathrm{PBH}=1$. (\textit{Upper}) The fiducial \PR model and variations all receive larger contributions from large halos, whereas the \BHL model's luminosity is dominated by small halos. This is due to the small DM-baryon relative velocity and ambient sound speed in smaller halos boosting the BHL accretion rate, while the PR accretion rate is limited by the high gas temperature due to feedback. (\textit{Lower}) When convolved with HMF, we see that the PR model receives the largest contribution from halos around $10^8\Msun$. At this mass, the BHL rate crosses over the PR rate and grows for smaller halos while the PR rate sharply decreases. This makes the BHL model much more susceptible to uncertainties on the small scale matter power spectrum.
    \label{fig:pbhacc-L-M}}
\end{figure}

We assume that PBH act as test masses that fully virialize within their host halos, and we use a Jeans analysis to evaluate the velocity distribution associated with a smooth NFW profile in order to determine the PBH-baryon relative velocities. As far as the validity of this approximation, it is likely a good one in the case of PR accretion for the reason that most of the accretion-powered luminosity is generated by PBHs which reside in halos of masses  $M_\mathrm{halo} \gtrsim 10^7 \, M_\odot$ while we consider PBH masses between $1$ and $10^{4}$ $M_\odot$, representing a large hierarchy. In Fig.~\ref{fig:pbhacc-L-M}, we validate this argument by presenting the dependence of the accretion luminosity of a $100\,\mathrm{M}_\odot$ PBH as a function of its host halo mass and the differential accretion luminosity per host halo mass generated by convolving the PBH luminosity as a function of host halo mass with the halo mass function, assuming $f_\mathrm{PBH} = 1$. 

On the other hand, in the BHL model, we find that the very low velocity dispersions associated with very light host halos causes the dominant fraction of the accretion luminosity to come from host halos very near in mass to the PBH. In this work, we examine the parametric dependence of our projected sensitivities as a function of the minimum mass of PBH-hosting halo within the BHL model, which we systematically vary from a minimum halo mass of 30 times that of the PBH mass in our fiducial BHL scenario to a minimum halo mass of 100 times the PBH mass in the \BHLmt scenario. This larger minimum halo mass will result in reduced projected sensitivity, though this procedure is certainly an ad hoc one. If BHL is indeed an accurate description of PBH accretion, then a more detailed understanding of the formation, survival, and possible disruption of halos host to comparable mass PBHs is critical for more robust sensitivity estimates and searches in data.

\begin{figure}[ht!]
    \centering
    \includegraphics[width=0.99\linewidth]{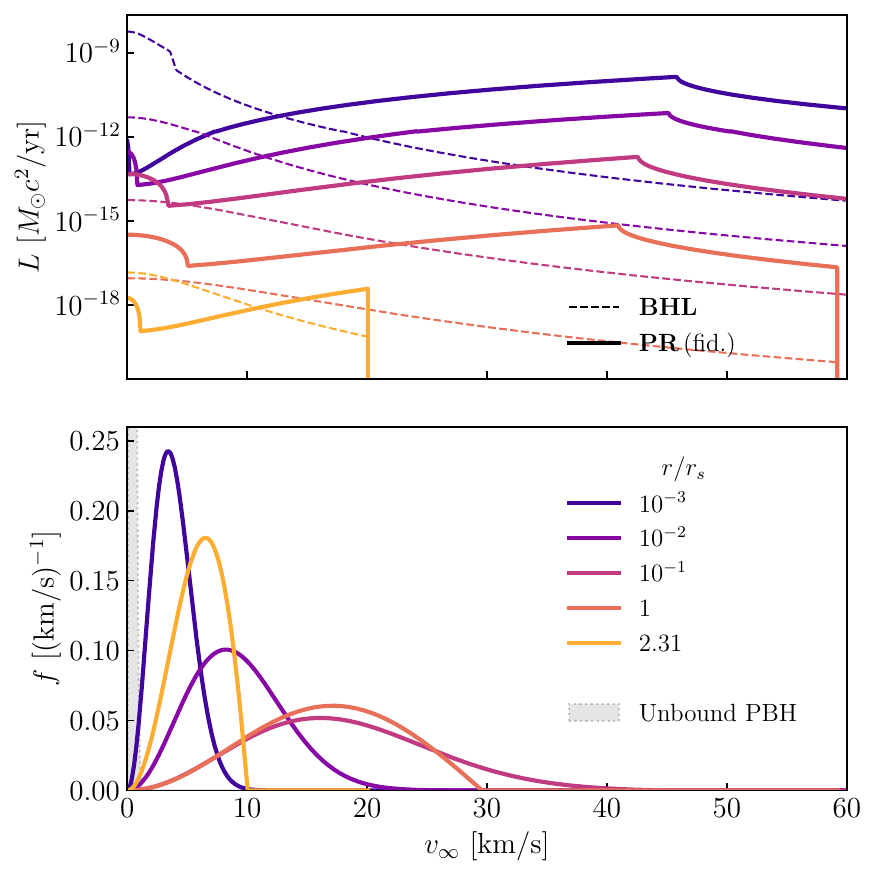}
    \caption{\textbf{Example single PBH accretion luminosity $L(v_\mathrm{rel},r)$ and relative velocity distribution $f(v_\mathrm{rel})$ in a $10^8\Msun$ halo.} The PBH mass is taken to be $M_\mathrm{PBH}=100\Msun$, with $f_\text{PBH}=1$. This snapshot is taken at redshift $z=15$, the halo concentration is $c=2.57$ with scale radius $r=351$~pc. $L$ values for the \PR and the \BHL models, and $f$ values at various radii are shown. As $r$ decreases towards small values, the expected luminosity increases due to higher baryon concentration and decreased relative velocity distribution.
    }
    \label{fig:pbhacc-halo-L-f-v}
\end{figure}

Our Jeans analysis, which predicts a radially dependent velocity distribution as a function of the halo density profile, is described in App.~\ref{app:f-v-rel}. As an example, for a $10^8\,\Msun$ halo with a concentration of $c = 2.57$, we illustrate the inferred velocity distribution for several radii within the halo in the bottom panel of Fig.~\ref{fig:pbhacc-halo-L-f-v}. In the top panel, we illustrate the accretion luminosity as a function of the PBH velocity for each of those radii, calculated in our fiducial \PR accretion model.

A necessary input for our accretion luminosity calculation is the PBH-gas relative velocity, and while the Jeans analysis allows us to predict the PBH velocity, modeling the phase space of baryons is more challenging as it requires accurately modeling the role of their nongravitational interactions across orders of magnitude in host halo mass.\footnote{These baryonic interactions may also spoil the assumption we have made of a spatially uniform ratio of baryonic matter to dark matter in the host halo.} In order to predict the accretion luminosity, we then consider two modeling assumptions to predict the gas bulk velocity on the $d\lesssim 1\,\mathrm{pc}$ scale corresponding to the size of the Str\"omgren sphere range for astrophysical BHs \cite{PR13, PR13I, PR13II}. 

In the first scenario, which we incorporate in our fiducial model, we take the bulk velocity of gas to be zero such that the PBH-baryon relative velocity merely follows the Jeans analysis velocity distribution. We also take the sound speed in the gas to be
\begin{equation}
    c_\infty(r)=\sqrt\frac{5}{9}\,v_0(r),
\end{equation}
corresponding to the relation between the velocity dispersion parameter $v_0(r)$ of a Maxwell-Boltzmann distributed ideal gas and its sound speed. The expected luminosity of a PBH located at a radius $r$ within a host halo is then given by
\begin{equation}
    \langle L_\mathrm{PBH} \rangle (r) = \int d^3\mathbf{v}_\infty f(\mathbf{v}_\infty |r) L(\rho_b(r),v_\infty(r), c_\infty(r))
\end{equation}
where the dependence on $M_\mathrm{PBH}$ and the host mass and density profile is left implicit. In the second scenario, which we designate by \PRB, we instead assume that the gas bulk velocity and PBH velocity both follow the Jeans analysis velocity distribution so that 
\begin{equation}
    \langle L_\mathrm{PBH} \rangle (r) = \int d^3\mathbf{v}_\infty f_\mathrm{rel}(\mathbf{v}_\infty |r) L(\rho_b(r),v_\infty(r), c_\infty(r)),
\end{equation}
with a definition of $f_\mathrm{rel}$ provided in App.~\ref{app:f-v-rel}. We consider these two scenarios to bracket our uncertainty on the gas bulk velocity. The total expected luminosity from a halo of mass $M$ is then
\begin{equation}
    \langle L_\text{halo}\rangle(M, z)= \frac{\Omega_\mathrm{DM} f_\mathrm{PBH}}{\Omega_\mathrm{DM}+\Omega_\mathrm{b}}\int d^3\mathbf{r} \frac{\rho_\mathrm{NFW}(\mathbf{r})}{\Mpbh}\langle L_\text{PBH} \rangle (\mathbf{r}).
\end{equation}
where  $\rho_\mathrm{NFW}$ is the NFW profile describing the host halo and the  $f_\mathrm{PBH}$ is, as before, the relative abundance of PBHs with respect  to the DM abundance.
Integrating over a halo mass function, we then arrive at the luminosity density in one cell of our simulation.

In both cases, we can calculate the local rate of energy per unit volume injected by halo-bound PBH accretion by
\begin{equation}
    \frac{dE}{dV dt}(z, x)=\int_{\Mpbh}^\infty dM\frac{dN}{dM}(\delta(x),z)\langle L_\text{halo}\rangle(M, z),
\end{equation}
where $dN/dM$ is the local conditional halo mass function, which is calculated according to App.~\ref{App:EPS}. While we discuss our model variations in greater length in Sec.~\ref{sec:BHModelVariationsResults}, we generally find that the treatment of the gas velocity has a negligible impact on the predicted luminosity, which is generally unsurprising as the two cases correspond to only an order unity variation in the PBH-baryon relative velocity.

As a modeling choice, we integrate only over halos with mass greater than a threshold  {$M^\text{low}_\mathrm{th}$ that is 30 times the PBH mass (or the sum with the UCMH mass for the variant \PRH), as PBHs will disrupt the formation of comparable mass halos. This choice represents an ad-hoc modeling choice that has limited impact on the projected sensitivities under PR-type models, but has considerable impact on projections developed under the BHL model, which we discuss at greater length in Sec.~\ref{sec:BHModelVariationsResults}. In Fig.~\ref{fig:pbhacc-L-M}, we illustrate the single halo luminosity as a function of mass, and the convolution of the single halo luminosity with the HMF; further discussion of the model variations is presented in Sec.~\ref{sec:BHModelVariationsResults}. Consistent with \cmfast, we compute the luminosity density for each cell using the Press-Schechter conditional halo mass function (HMF).

\begin{figure}[!t]
    \centering
    \includegraphics[width=\linewidth]{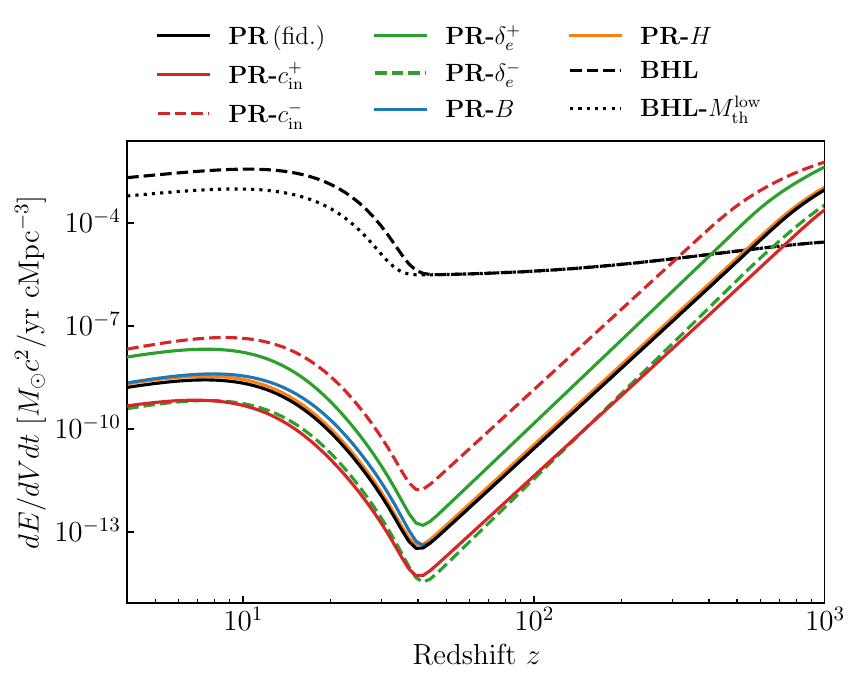}
    \caption{\textbf{PBH luminosity density history for various accretion models for $M_\mathrm{PBH}=100\Msun$, $f_\mathrm{PBH}=1$.} We compare the spatially averaged luminosity history of our fiducial \PR accretion model with variations, including the \BHL models. All models transition from being dominated by unbound PBHs to halo PBHs, with PR models at around $z=45$, and BHL models earlier as they receives more contribution from small halos. The PR model variations are mostly bound by the \PRcm and \PRcp models, in which the interior sound speed of the accretion region is changed. The model \PRB and \PRH almost overlap with the fiducial PR model. See text for more details on model variations.}
    \label{fig:pbhacc-L-z}
\end{figure}

Let us now make a direct comparison to the unbound PBH accretion luminosity studied in Sec.~\ref{sec:unbound_pbh}. As before, we take $z=15$, and we now consider a PBH in a host halo of mass $10^8\,M_\mathrm{\odot}$; from Fig.~\ref{fig:pbhacc-L-M}, we see that in all PR model variations, this is the host halo mass scale associated with the majority of bound PBH accretion luminosity. The \texttt{Luldow16} concentration mass relation predicts a concentration of $c=2.57$, corresponding to a scale radius $r_s=352$~pc. For a PBH at a radius of $r_s/10$, assuming an NFW distributed gas profile, the local gas density is 91/cm$^3$, much higher than the cosmological average, the RMS relative velocity is $v_\mathrm{rel}$ is 19.7~km/s, and the ambient gas sound speed is 12~km/s. Assuming our \PR fiducial model with $\cin=23$~km/s, the PBH sits in the intermediate velocity regime. For a $1~\Msun$ PBH, the accretion rate is $9.6\times10^{-11}\Msun/$yr, and the total emission luminosity is at $2.6\times10^{34}$~erg/s, roughly 13 orders of magnitude larger than the a unbound PBH accretion rate.\footnote{We have chosen the benchmark these illustrative values for a PBH displaced from the center of the halo by a radius of $r = r_s/10$ because the distribution of the total PBH accretion luminosity over the halo is maximized at around this radius. As a result, this radius is the most representative one for PBHs that make the dominant contribution to the total accretion-powered luminosity and lead to our projected sensitivities.}

Moreover, in Fig~\ref{fig:pbhacc-L-z}, we illustrate the PBH energy injection rate $dE/dVdt$ as a function of redshift, calculated assuming a local overdensity of $\delta=0$ for each of the model variations in this work. While the PR model variations and the BHL model predict quantitatively different energy injection histories, they are qualitatively similar in that turnover from energy injection being dominated by unbound PBH accretion to being dominated by halo-bound PBH accretion occurs at $z\approx 50\sim100$.

\subsubsection{Ultracompact Minihalos}
\label{sec:UCMHs}

It has been argued that the radial, self-similar infall of matter onto a PBH or other compact object would produce dense UCMHs. The formation and associated density profile of these UCMHs have been studied extensively, though with mixed conclusions \cite{MOR06, Berezinsky:2013fxa, Eroshenko:2016yve, Delos:2017thv, Adamek:2019gns, Boudaud:2021irr}. If these UCMHs do indeed form, then they could impact the accretion rate and luminosity of PBHs. We consider here as a model variation the impact of UCMHs that may form around PBHs, which would affect both unbound and halo-bound PBHs. 

We realize the effect of UCMHs in \dmcm through a modification to the environmental dependence of the black hole accretion rate and luminosity. We assume that if UCMHs do form, the baryonic abundance in the UCMH is that of the large-scale cosmological average, and we implement an analytic UCMH model closely following the prescription of Ref.~\cite{AEGSSV24} used to study the impact of UCMHs on PBH accretion in the context of the CMB. This modeling prescription is taken to replace the accretion rate and luminosity calculation of both unbound and halo-bound accreting PBHs alike.

The PR accretion rate of Eq.~\ref{eq:PR-acc-rate} can be rewritten as 
\begin{equation}
    \dot M_\text{PR}=4\pi\rho_\mathrm{in}v_\mathrm{eff}\left(r_\mathrm{B}\right)^2,
\end{equation}
where $v_\text{eff}=\sqrt{v_\text{in}^2+c_\text{in}^2}$, and $r_\mathrm{B}=GM/v_\text{eff}^2$ is the usual Bondi radius, characterizing the gas-capturing cross section. In the presence of UCMHs, this radius is replaced with an increased, effective Bondi radius $r_\mathrm{B,eff}$ that satisfies the following relation
\begin{equation} \label{eq:veff-mod}
    v^2_\mathrm{eff}=\frac{GM}{r_\mathrm{B,eff}}-\phi_h(r_\mathrm{B,eff}),
\end{equation}
where $\phi_h$ is the additional gravitational potential generated by the UCMH \cite{SPIK20}. The effective Bondi radius is then solved for given $v_\mathrm{eff}$. Hence, given a UCMH density profile, which also specifies the baryonic density, we can evaluate an updated accretion rate. Note that as before, we have assumed that the capture happens entirely within the ionized region considered in the PR model, consistent with and supported by the arguments of Ref.~\cite{AEGSSV24} and Ref.~\cite{SR20}.
Following \cite{Bertschinger:1985pd}, we model the UCMH profile by
\begin{equation}
    \rho_\mathrm{UCMH} \propto r^{-9/4},
\end{equation}
and take the radius and total mass to be redshift-dependent and parametrized by
\begin{equation}
    M_h=\frac{3000}{1+z}\Mpbh,\quad r_h=\frac{58\mathrm{ pc}}{1+z}\left(\frac{M_h}{M_\odot}\right)^{1/3},
\end{equation}
as in Ref.~\cite{SPIK20}, though we cutoff the mass growth if $f_\mathrm{PBH} M_h/M_\mathrm{PBH}$ exceeds $1$, which corresponds to overclosure by the halos. In total, we find that inclusion of the possible UCMH enhances the accretion rate by an $\mathcal{O}(1\sim 10)$ amount, consistent with the findings at earlier CMB-relevant times of Ref.~\cite{AEGSSV24}. We denote the model variation associated with including the presence of UCMHs by \PRH.

\subsection{Model Variations and Results}
\label{sec:BHModelVariationsResults}
After describing a lengthy modeling effort, we pause here to briefly summarize the fiducial model and modeling variations before presenting our projected sensitivity to accreting PBHs.

In our fiducial model, we assume a Park-Ricotti type accretion model \cite{PR13, PR13I, PR13II} with a interior sound speed in the ionized region of accretion $c_\mathrm{s, in}= 23\,\mathrm{km}/s$ with an efficiency factor relating accretion modeled according to \cite{XY12} assuming an electron-ion coupling of $\delta_e = 0.1$. We self-consistently account for accretion-powered luminosity from PBHs that are unbound to halos and those that are bound to halos, with the bound halo population modeled assuming they virialize into host halos following a mass function computed with the extended Press Schechter formalism that are also host to gas with negligible bulk velocities compared to the PBH peculiar velocities. We do not consider the formation of ultracompact minihalos or other dense structures or substructures in the vicinity of the PBHs. The accretion luminosity predictions associated with this fiducial model are illustrated in black in Fig.~\ref{fig:pbhacc-L-M} and Fig.~\ref{fig:pbhacc-L-z}, and it is used to develop the luminosity predictions in Fig.~\ref{fig:pbhacc-halo-L-f-v}.

To compare modeling variations within the PR modeling paradigm, we bracket the uncertainties on the ionized region sound size by varying otherwise fixed fiducial model to use $c_\mathrm{s,in}= 14\,\mathrm{km/s}$ and $c_\mathrm{s,in}= 29\,\mathrm{km/s}$ in the \PRcm and \PRcp scenarios, respectively. Likewise, we vary the electron-ion coupling which determines the luminosity efficiency factor to $\delta_e = 0.01$ and $\delta_e= 0.5$ in the \PRdm and \PRdp scenarios, respectively. While our study now includes the effect of accretion in halo environments, here too we find that the uncertainty in the sound speed represents the most important uncertainty within the PR modeling paradigm, with our bracketing range corresponding to an order of magnitude variation in the predicted energy injection rate across cosmological times. The uncertainty associated with the electron-ion coupling is nearly as significant in terms of its effect on the energy injection rate.

By contrast, our systematic variations which seek to understand the dependence of PBH accretion on the halo environment reveal that our predictions are relatively robust to the modeling of the halo structure. Considering the model variation \PRB, which took the gas velocity to follow the Maxwell-Boltzmann velocity distribution for gravitationally virialized matter, and model variation \PRH, which considered the impact of UCMHs which form around PBHs, only resulted in order unity changes in the predicted energy injection rate. 

However, the most significant model variation is the choice to use the BHL accretion model rather than a PR accretion model. As seen in Fig.~\ref{fig:pbhacc-L-z}, even with our $\lambda = 10^{-2}$ phenomenological suppression factor, the BHL accretion model predicts many orders of magnitude greater rates of energy injection from PBH accretion at $z\lesssim 300$, and Fig.~\ref{fig:pbhacc-L-M} reveals this results from a boosted accretion luminosity associated with the lightest halos which host PBHs. Physically, this is because the factor of $1/(v^2 + c_s^2)^{3/2}$ which appears in the BHL accretion rate diverges with decreasing halo mass while the analogous factor in the PR model is regulated by the induced gas velocity and sound speed in the ionized region. It is likely that the BHL accretion rate predictions in this context are significant overestimates since, even neglecting the baryonic dynamics, the massive PBH is likely to gravitationally disrupt or backreact on the host halo structure. We emphasize that this is necessarily not a fundamental problem of the BHL accretion model; if the BHL model accretion is indeed an accurate description for PBH accretion, then this indicates the importance of improved modeling of the dynamics of accreting BHs when their mass is no longer negligible compared to the mass of their host halo. 

\begin{figure}[!ht]
    \begin{center}
    \includegraphics[width=\linewidth]{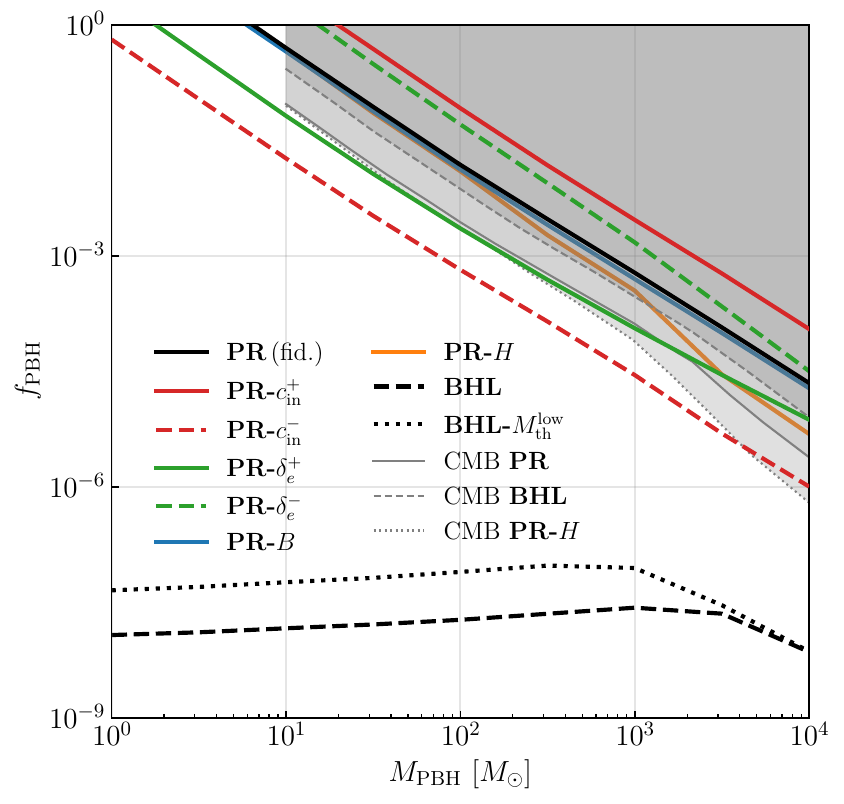}
    \caption{\textbf{Projected 95$^\text{th}$ percentile upper limits on accreting PBHs.} The \PR model limits are comparable with CMB limits, while theoretical uncertainties remain large. The \PR limits are most significantly affected by the sound speed in the ionized region $c_\mathrm{in}$, and secondly the ADAF electron heating fraction $\delta_e$. Effects of baryon velocity distribution and UCMH (\PRB and \PRH respectively) remain small except for UCMH at high PBH masses. The dashed and dotted black lines represent the fiducial \BHL model, and a variation \BHLmt where the small halo mass threshold is raised by a factor of 3.3, which significantly impacted the sensitivity. At low PBH masses in the \BHL model, the sensitivity is only weakly dependent on the PBH mass since decreasingly massive PBHs may reside within decreasingly massive host halos, with yet smaller velocity dispersions and greater accretion luminosities. The CMB limits are quoted from \cite{AEGSSV24}.}
    \label{fig:pbhacc-limits}
    \end{center}
\end{figure}

For each of these modeling variations, we apply our simulation and projected sensitivity formalism described in Sec.~\ref{sec:review} to develop projected sensitivities on $f_\mathrm{PBH}$ as a function of PBH mass assuming monochromatic PBH mass functions. Those results are presented in Fig.~\ref{fig:pbhacc-limits} alongside existing limits on the PBH landscape. Since PR-type accretion loses sensitivity at $M_\mathrm{PBH}\sim 1\,\mathrm{M}_\odot$, we do not consider lower masses, though for BHL-type accretion, sensitivity would extend down to PBH masses comparable to the minimum halo mass, which is generally DM- and inflationary-model dependent. 

Moreover, despite that PR-type accretion sensitivity improves with increasing PBH mass, we do not extend our projected reach to higher ranges for two reasons. First, more massive PBHs would become comparable in mass to the largest possible host halos at times relevant for reionization, requiring a more careful treatment of host halo modeling. Moreover, the projected constraints on $f_\mathrm{PBH}$ become sufficiently strong that the expected number of PBHs per lattice volume in our simulations would become less than $1$ such that shot-noise contributions to the 21-cm power spectrum may become important. Understanding the behavior of large PBHs in small host halos would be particularly valuable as it would also lead to a better understanding of model predictions in the BHL paradigm. Studies of these matters are beyond the scope of the work we present here, but represent interesting directions for future study.

It is particularly interesting that BHL accretion results in greatly enhanced sensitivity to PBH accretion through 21-cm probes. By contrast, in the context of the CMB, BHL accretion is associated with somewhat reduced PBH sensitivity as compared to PR accretion, though this hierarchy is reversed under the assumption of UCMHs. \cite{AEGSSV24}. In fact, while better sensitivity in PR-type models is realized by the CMB, better sensitivity in BHL-type models is realized by 21-cm. As a result, probing PBH accretion in the context of 21-cm and the CMB offers an opportunity to bracket the accretion modeling uncertainty with future work that more precisely characterizes the efficiency and luminosity of PBH accretion, enabling an identification of the optimal cosmological probe.

\section{p-Wave Annihilating Dark Matter}
\label{sec:dm_annihilation}
The last of the three scenarios that we consider in this work is that of $p$-wave annihilation. As a brief review, if cold DM does annihilate, its annihilation cross-section can be written order-by-order in the relative velocity of the two particles by 
\begin{equation}
    \langle\sigma v\rangle = \langle\sigma v\rangle_s + C_\sigma\frac{\langle v_\mathrm{rel}^2 \rangle}{c^2}+\cdots.
\end{equation}
Here, $\langle \rangle$ indicates an averaging over the DM velocity distribution. The first term in this expansion $\langle \sigma v \rangle_s$ represents the velocity-averaged \textit{s}-wave cross section, which is, in many models, the leading order contribution and is velocity-independent. However, symmetry considerations may forbid the \textit{s}-wave contribution, resulting in a leading contribution from \textit{p}-wave annihilation which scale with the square of the relative velocity of the DM. In this work, we consider \textit{p}-wave annihilation in a model-independent manner, with the total \textit{p}-wave annihilation cross-section set to the characteristic cross section $C_\sigma$. Some specific model constructions in which \textit{p}-wave processes are the leading contribution to DM annihilations include fermionic Higgs portal DM \cite{Kim:2006af, Kim:2008pp} and charged scalar DM that annihilates through an \textit{s}-channel gauge boson \cite{Hagelin:1984wv}. For more general model constructions of \textit{p}-wave annihilating DM, see \cite{Pospelov:2007mp, Evans:2017kti}.

Because \textit{p}-wave annihilation is velocity-suppressed, it is generally challenging to detect, and it realizes a considerably different environmental dependence than \textit{s}-wave annihilation. Because the \textit{s}-wave annihilation is velocity-independent, it depends strongly on the small-scale structure of DM, since the smallest halos, although characterized by small virial velocities, are typically denser than larger, later-forming halos. By contrast, the \pwave annihilation rate is suppressed in small scale structures due to their low virial velocity, and the total rate is dominated by contributions from larger halos realized at later times, as shown in Fig.~\ref{fig:pwave-hmf-contribution}. As a result, 21-cm probes provide an opportunity to improve upon previous exotic energy injection constraints based on the CMB \cite{Diamanti:2013bia, Wang:2025tdx}.

\subsection{\textit{p}-wave Implementation in \dmcm}

To realize \textit{p}-wave annihilation in the \dmcm modeling framework, we express the energy injection  as in Eq.~\ref{eq:EnergyInjectionRate} by 
\begin{equation}
\begin{split}
    \frac{dN_{\gamma/e}}{dE dV dt} = C_\sigma  \frac{dN_{\gamma/e}}{dE}\int dM &\frac{dN}{dM}(z, \mathbf{x}) \mathcal{L}(M,z)
\end{split}
\end{equation}
where $dN/dM$ is the local halo mass function calculated according to App.~\ref{App:EPS}, $\mathcal{L}$ is the annihilation luminosity of a halo of mass $M$ at time $t$ per unit $C_\sigma$, and $dN_{\gamma/e}/dt$ is, as before, the unit-normalized spectrum of secondary photons/electrons produced in an annihilation event. In this work, we consider only field halos as the annihilation boost from subhalos for \pwave annihilation is expected to be unimportant due to their smaller mass and velocity dispersion. We now specify our calculation of the local halo mass function and the halo annihilation luminosity.

To determine the halo annihilation luminosity, we take halos to follow NFW profiles specified by an enclosed mass $M_{200}$ and a concentration parameter $c_{200}$ \cite{Navarro:1996gj, 2010gfe..book.....M}. As before, we model the redshift-dependent concentration-mass relationship $c_{200}(M_{200}, z)$ with the \texttt{Ludlow16} model \cite{Ludlow:2016ifl, Murray:2013qza,Murray:2020dcd}. The total annihilation rate in the halo is then given by 
\begin{equation}
\begin{split}
\Gamma_a &= \frac{1}{2} \int dV \langle \sigma v\rangle n_\mathrm{DM}^2\\
&= \frac{C_\sigma}{2} \left(\frac{\Omega_\mathrm{DM}}{\Omega_\mathrm{DM} + \Omega_\mathrm{b}} \right)^2 \int dV \langle v_\mathrm{rel}^2(r) \rangle  \rho(r)^2.
\end{split}
\end{equation}
where $\langle v_\mathrm{rel}^2 \rangle(r)$ is the expected square of the relative velocity of two DM particles in the halo as a function of radius. For an NFW halo of mass $M$ and concentration $c$, this quantity can be evaluated efficiently under a Maxwell-Boltzmann approximation, see  App.~\ref{app:f-v-rel}. The luminosity is then straightforwardly given
\begin{equation}
    \mathcal{L}(M, z) = 2 m_\chi \Gamma_a
\end{equation}
where $m_\chi$ is the mass of the annihilating DM. 

Our calculation of the halo mass function has the advantage of accounting for spatial inhomogeneity through its conditioning on the local overdensity evaluated in the \dmcm simulation. However, as we discuss at greater length in App.~\ref{App:EPS}, our treatment cannot account for halos which are larger than the total mass residing within a simulation cell. For our simulations, which are performed with a comoving resolution of $2\,\mathrm{Mpc}$, this corresponds to a grid-scale mass of $M_\mathrm{grid} \approx 3 \times10^{11}\, M_\odot$. Since \textit{p}-wave annihilation is more efficient in more massive halos with correspondingly larger velocity dispersions, it is useful to examine the relative contributions to the total expected annihilation luminosity from halos with masses $M \gtrsim M_\mathrm{grid}$. To do so, we consider the halo mass function $dN/dM$ calculated for $\delta = 0$ as a function of time, and, more saliently, the fractional cumulative distribution of \textit{p}-wave annihilation by halo mass given by
\begin{equation}
    F_{\mathcal L}(M,z) = \frac{\int^{M} dM'\frac{dN}{dM'}(\delta = 0,z) \mathcal{L}(M',z)}{\int dM'\frac{dN}{dM'}(\delta = 0,z) \mathcal{L}(M',z)}.
\label{eq:FractionalCDF}
\end{equation}
If $F_{\mathcal L}(M_\mathrm{grid},z) \ll 1$, then it indicates that most of the \textit{p}-wave annihilation is expected to occur in halos above our grid-scale mass, rendering our EPS treatment insufficient. However, as we demonstrate in Fig.~\ref{fig:pwave-hmf-contribution}, this is not the case, and the vast majority of \textit{p}-wave annihilation occurring at any time occurs within halos below our grid-scale mass and hence is captured by our EPS treatment. As further evidence that our projected sensitivities are robust with respect to this resolution limitation, in Sec.~\ref{sec:pwave_results}, we will also develop projected sensitivities imposing an artificial high cutoff to our halo mass function at $M_\text{th}^\text{high} = M_\mathrm{grid}/3$.

\begin{figure}[h!]
    \centering
    \includegraphics[width=\linewidth]{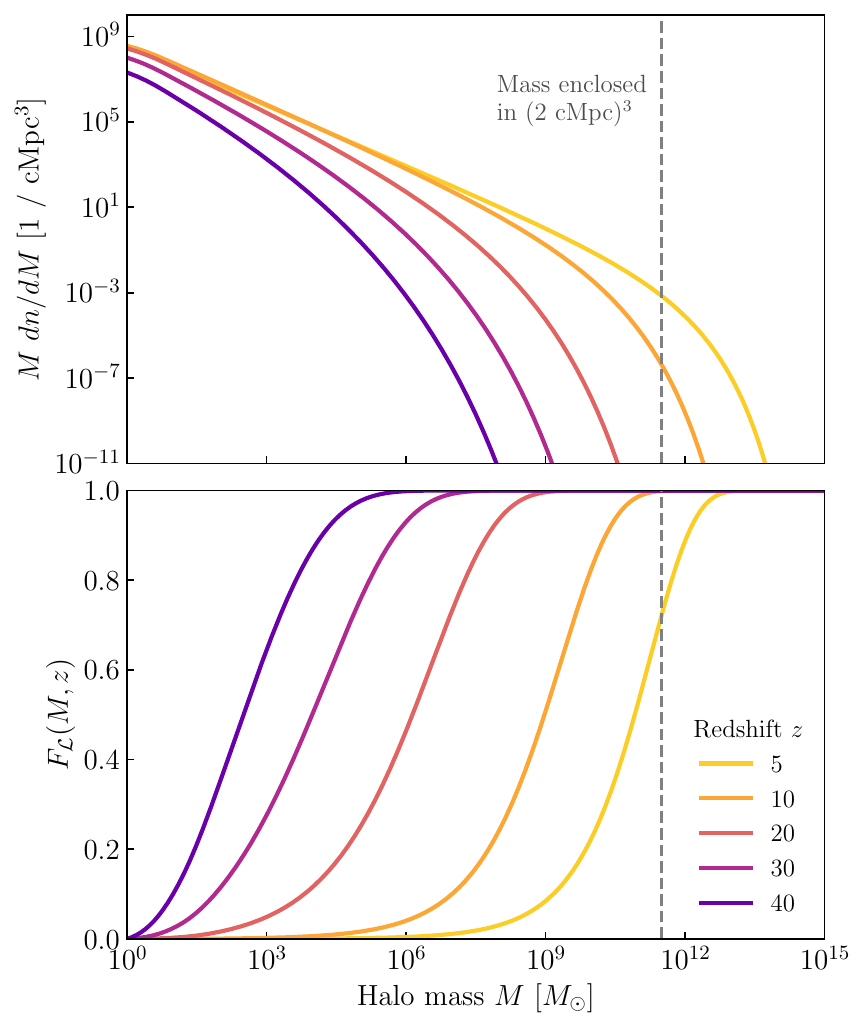}
    \caption{ \textbf{Halo mass function and \textit{p}-wave annihilation contributions by redshift.} In the top panel, we illustrate the halo mass function at a range of redshifts between $z = 5$ and $ z = 40$, showing that even at redshifts as low as $z = 5$, the number of halos with mass greater than our grid-scale mass (indicated by the vertical gray dashed line) is relatively small. In the bottom panel, we show the fractional cumulative distribution function (CDF) for annihilation luminosity (see Eq.~\ref{eq:FractionalCDF}). The intersection of the dashed gray line corresponding to the grid-scale mass and the CDF indicates the fraction of total \textit{p}-wave annihilation luminosity within halos less than the grid-scale mass that are within the resolution of our halo mass function subgrid modeling described. Even at the latest redshift, $z = 5$, relevant for our analysis, only $\sim$25\% of annihilation is expected to occur in halos larger than our grid-scale mass. At earlier times, neglecting halos with mass below the grid-scale mass is an even better approximation.}
    \label{fig:pwave-hmf-contribution}
\end{figure}

\subsection{Results}
\label{sec:pwave_results}
As before, following the methodology developed in Sec.~\ref{sec:review}, we compute 21-cm power spectra in the presence of \textit{p}-wave annihilation considering three benchmark scenarios: annihilation to photons, annihilation to electrons, and annihilation to taus. For each scenario, we consider DM masses as low as the maximum of $10^{2}\,\mathrm{eV}$ and the smallest kinematically allowed mass for the given annihilation channel and as high as $10^{12}\,\mathrm{eV}$. We develop projected sensitivities on the \textit{p}-wave annihilation cross-section parameter $C_\sigma$, presented for each of our three annihilation channels of interest in Fig.~\ref{fig:pwave-limits}. In addition to existing limits on $p$-wave annihilation, we also convert $s$-wave limits by setting $\langle v_\text{rel}^2\rangle = (220\text{~km/s})^{2}$ for limits made within the Milky Way halo, and $\langle v_\text{rel}^2\rangle = (1000\text{~km/s})^{2}$ for galaxy clusters used in \cite{Lisanti:2017qlb} while removing a \swave boost factor of 5. Results for additional annihilation channels are made publicly available in our code release. 

\begin{figure*}[!t]
    \begin{center}
    \includegraphics[width=\textwidth]{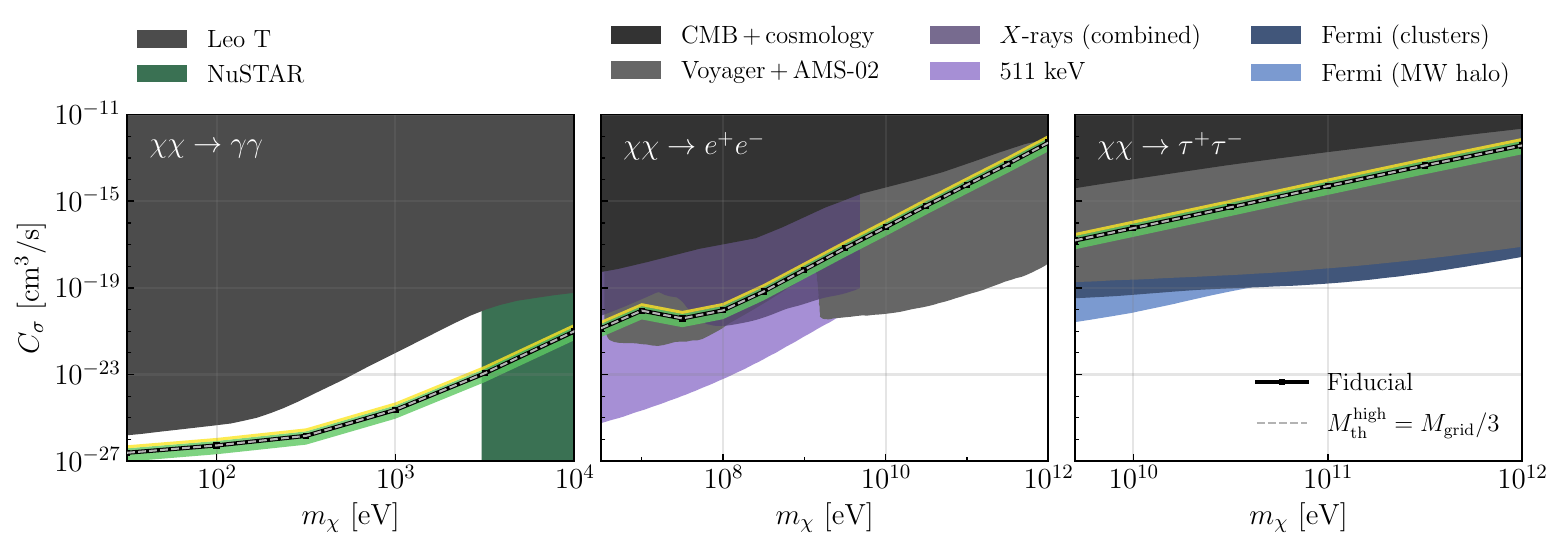}
    \caption{\textbf{Projected 95$^\text{th}$ percentile upper limits on \pwave annihilation.} The solid black lines represent our fiducial limits, while the dashed gray lines on top of the solid lines represent limits with $M_\text{th}^\text{high}=M_\text{grid}/3$, showing that our limits are not sensitive to $M_\text{th}^\text{high}$. Additional $p$-wave constraints (\textit{uncolored}) on the parameter spaces are provided by the heating of Leo T dwarf galaxy \cite{Wadekar:2021qae}, a joint analysis of CMB and several cosmological datasets \cite{Diamanti:2013bia}, and Voyager and AMS-02 data \cite{Boudaud:2018oya}. We also convert $s$-wave constraints (\textit{colored}) from NuSTAR \cite{Zakharov:2025coj}, a combined analysis of \xray data \cite{Cirelli:2023tnx}, 511~keV line \cite{DelaTorreLuque:2023cef}, Fermi observation of the Milky Way halo \cite{Chang:2018bpt}, and Fermi observation of galaxy clusters \cite{Lisanti:2017qlb}. See text for details.
    }
    \label{fig:pwave-limits}
    \end{center}
\end{figure*}

We also use these limits to examine the dependence of our projected sensitivities on the artificial cutoff imposed by our characteristic grid scale in the conditional halo mass function. Imposing a maximum mass of $M_\text{th}^\text{high}= M_\text{grid}/3$ on the halo mass function, we otherwise repeat an identical limit-setting procedure to develop the projected sensitivities shown in dashed black in Fig.~\ref{fig:pwave-limits}, demonstrating that our sensitivities are not strongly affected by this modeling limitation.

\section{Conclusion}
\label{sec:conclusion}

In this work we have extended the \texttt{DM21cm} code for modeling general, spatially inhomogeneous energy injection histories and applied it to three motivated new-physics scenarios: Hawking radiation from evaporating primordial black holes, accretion-powered emission from more massive PBHs, and velocity-suppressed (\pwave) dark matter annihilation. In addition, these scenarios are physically motivated ones of interest to the next generation of cosmic surveys; they also highlight qualitatively different modeling challenges for simulating the imprints of new physics on the 21-cm power spectrum. Decaying PBHs realize time-varying evaporation rates and spectra, accreting PBHs realize strongly environmentally dependent accretion-powered luminosities, and \pwave annihilation depends on the process of cosmological structure-formation.  By consistently incorporating these processes into simulations of the 21-cm power spectrum, we have demonstrated that near-term interferometric observations such as HERA can reach leading, or near-leading, sensitivity to each of these scenarios. 

The results presented here underscore the unique role of 21-cm cosmology as a probe of exotic energy injection at redshifts intermediate between those best constrained by the CMB and those accessible to late-time astrophysical searches. While the precise reach depends on modeling assumptions, especially for PBH accretion, where theoretical uncertainties remain large, the distinct spatial and temporal morphology of the 21-cm power spectrum provides powerful discriminatory power that is complementary to other cosmological and astrophysical probes. 

Looking ahead, further refinements to the modeling of PBH accretion, the treatment of small-scale structure, and joint analyses incorporating global 21-cm measurements, higher-point statistics, or cross-correlations with other probes will sharpen the reach of 21-cm cosmology. The public release of our extended \texttt{DM21cm} framework provides a flexible platform for carrying out such studies, enabling exploration of a broad landscape of new-physics scenarios.

\textbf{Note Added:} In the final stages of preparation of this manuscript, Ref.~\cite{Zhao:2025ekx} appeared, projecting sensitivity to Hawking radiation from PBH decay through 21-cm probes. This work differs from that which we present here in that it implements a less detailed energy injection modeling procedure and studies sensitivity from upcoming SKA measurements, rather than the more near-term HERA projections we develop.

\begin{acknowledgments}
We thank Hongwan Liu and Tracy Slatyer for helpful discussions and collaboration at early stages of this work. We thank Dominic Agius, Laura Lopez Honorez, Paolo Panci, and Mauro Valli for helpful conversations. YS was supported by a Trottier Space Institute Fellowship. JBM was supported by NSF through grants AST-2307354 and AST-2408637. The computations in this paper were run on the Erebus machine at MIT and the FASRC Cannon cluster supported by the FAS Division of Science Research Computing Group at Harvard University.

This work made use of
\cmfast \cite{2011MNRAS.411..955M},
\texttt{21cmfish} \cite{Mason:2022obt},
\texttt{21cmSense} \cite{pober201621cmsense},
\texttt{DarkHistory} \cite{Liu:2019bbm},
\texttt{HaloMod} \cite{Murray:2013qza, Murray:2020dcd},
\texttt{JAX} \cite{jax2018github},
\texttt{NumPy} \cite{harris2020array},
\texttt{Scipy} \cite{virtanen2020scipy},
\texttt{Astropy} \cite{robitaille2013astropy},
\texttt{matplotlib} \cite{hunter2007matplotlib},
and all associated dependencies.
\end{acknowledgments}

\appendix

\section{Details of the Park-Ricotti Model}
\label{app:pr_model}
Here we describe the details of the PR model necessary to reproduce the calculations in this work. For a more detailed discussion of the PR model, we refer the readers to \cite{AEGSSV24} and original references of the PR model, \cite{PR13,PR13I,PR13II,SR20}.

The PR accretion model describes accretion dynamics in the presence of an ionized region in the vicinity of the black hole generated by baryonic feedback. Since the ionized region in the PR model is larger than the Bondi radius, the accretion rate is taken to be described by Eq.~\ref{eq:PR-acc-rate}, which, unlike the BHL accretion rate of Eq.~\ref{eq:BHL-acc-rate}, does not include a fudge factor but is given in terms of interior quantities defined within the ionized region. The three relevant quantities are the interior density $\rho_\mathrm{in}$, the BH relative velocity with respect to the baryons,  relative velocity $v_\mathrm{in}$, and the interior sound speed $c_\mathrm{s,in}$.

The interior sound speed $c_\mathrm{s,in}$ is determined by the detailed microphysics of the heating and cooling mechanisms of the baryon gas. As discussed in Sec.~\ref{sec:accretion-modeling}, we have fixed it to a fiducial value of 23~km/s, with some systematic variations considered. The remaining two interior quantities can be evaluated through the requirement of continuity of mass and momentum over the interior ionized/exterior unionized boundary. These continuity requirements are given by 
\begin{equation}
\begin{aligned}
    \rho_\mathrm{in} v_\mathrm{in}&=\rho_\infty v_\infty \\\rho_\mathrm{in}(v^2_\mathrm{in}+c^2_\mathrm{s,in})&=\rho_\infty(v^2_\infty+c_{\mathrm{s},\infty}^2),
\end{aligned}
\end{equation}
enabling us to solve for the interior velocity and density as a function of the environmental velocity $v_\infty$, density $\rho_\infty$, and sound speed $c_{s, \infty}$, respectively. 

At low relative velocities $v_\infty < v_D\approx c_{\mathrm{s},\infty}^2/2c_\mathrm{s,in}$, a dense type (D-type) ionization front forms, corresponding to the solution
\begin{equation}
    \rho_\mathrm{in}=\rho_\infty\frac{v_\infty^2+c_{\mathrm{s},\infty}^2-\sqrt{\Delta}}{2c^2_\mathrm{s,in}},\qquad v_\mathrm{in}=v_\infty\frac{\rho_\infty}{\rho_\mathrm{in}},
\end{equation}
where the determinant is
\begin{equation}
    \Delta=(v_\infty^2+c_{\mathrm{s},\infty}^2)^2 - 4v_\infty^2 c_\mathrm{s,in}^2.
\end{equation}
At high relative velocities $v_\infty > v_R\approx2c_\mathrm{s,in}$, a rarefied type (R-type) ionization front forms, corresponding to the solution
\begin{equation}
    \rho_\mathrm{in}=\rho_\infty\frac{v_\infty^2+c_{\mathrm{s},\infty}^2+\sqrt{\Delta}}{2c^2_\mathrm{s,in}},\qquad 
    v_\mathrm{in}=v_\infty\frac{\rho_\infty}{\rho_\mathrm{in}}.
\end{equation}
In the intermediate velocity range, a shock front forms in front of the ionized region, and $v_\mathrm{in}$ is observed to reach the sound speed $c_\mathrm{s,in}$ \cite{PR13}. We then obtain
\begin{equation}
    \rho_\mathrm{in}=\rho_\infty\frac{v_\infty^2+c_{\mathrm{s},\infty}^2}{2c^2_\mathrm{s,in}},\qquad v_\mathrm{in}=c_\mathrm{s,in}
\end{equation}
for the interior density and velocity.

\section{Halo Velocity Distributions}
\label{app:vDisp}
In this Appendix, we describe how a Jeans analysis, in concert with a Maxwell-Boltzmann approximation may be used to predict the local velocity distribution of a spherically symmetric halo. We follow the procedure described in \cite{Lacroix:2020lhn}, which cites accuracy for velocity-related observables consistent at the $\sim20\%$ level with simulation, representing a small systematic uncertainty in the context of this work.

First, assuming a spherical DM halo with an isotropic velocity distribution, the radially dependent velocity dispersion is given by 
\begin{equation}
    \langle v^2(r) \rangle= \frac{3 G}{\rho} \int_r^{r_\mathrm{max}} dr' \frac{\rho(r') m(r')}{r'^2}
\end{equation}
where $\rho(r)$ local density of the halo at $r$,$m(r)$ the enclosed mass within a radius $r$, and $r_\mathrm{max}$ the maximum radius of the halo. In this work, use the $M_\mathrm{200}$ mass definition with associated concentration parameter $c_\mathrm{200}$, with implicit maximum radius $r_\mathrm{200}$ \cite{2010gfe..book.....M}.

Subject then to the Maxwell-Boltzmann approximation, the radially-dependent velocity distribution is given by
\begin{equation}
f(\mathbf{v},r) \propto \left[1 - \exp\left( -\frac{v^2-v_e^2 }{v_0^2} \right) \right]
\label{eq:velocity_dist}
\end{equation}
with support on $|\mathbf{v}| \leq v_e(r)$ where $v_e$ is the escape velocity for a particle at radius $r$ in the halo and $v_0(r) = \sqrt{2 \langle v(r)^2\rangle/3}$. The constant of proportionality can then be defined by the condition $\int d\mathbf{v} f(\mathbf{v}) = 1$. Note that this velocity  distribution is radially dependent through the implicit radial dependence of $v_e$ and $v_0$.

\section{Efficient Relative Speed Expectations}
\label{app:f-v-rel}
In this Appendix, we show how expectation values of functionals of the relative speed may be efficiently calculated, building on the results of App.~\ref{app:vDisp}.

Consider a quantity $Q$ which depends only on the relative speed so that 
\begin{equation}
    Q(\mathbf{v}, \mathbf{w}) = Q(|\mathbf{v}-\mathbf{w}|).
\end{equation}
We can evaluate $\langle Q \rangle$ by computing 
\begin{equation}
    \langle Q \rangle = \int d^3\mathbf{v} d^3\mathbf{w} f(\mathbf{v}) f(\mathbf{w}) Q(|\mathbf{v} - \mathbf{w}|).
\end{equation}
This can be written in terms of the relative speed distribution $f_\mathrm{rel}$ as
\begin{equation}
\begin{gathered}
    \langle Q \rangle = \int du f_\mathrm{rel}(u) Q(u) \\
    f_\mathrm{rel}(u) = u^2 \int d\Omega_\mathbf{u}  d^3 \mathbf{v} \, f(\mathbf{v}) f(\mathbf{v}- \mathbf{u}).
\end{gathered}
\end{equation}
With a change of variables which exploits the isotropy of the velocity distribution, we have
\begin{equation}
    f_\mathrm{rel}(u) = 8 \pi^2 u \int_0^\infty dv \, v f(v) \int_{|v - u|}^{v + u} dw \, w f(w).
\end{equation}
It is then straightforward to substitute in the velocity distribution of Eq.~\ref{eq:velocity_dist} so that we obtain
\begin{equation}
    f_\mathrm{rel}(u) \propto 4 \pi u^2 F\left(\frac{u}{v_0}, \frac{v_e}{v_0}\right)
\end{equation}
with 
\begin{equation}
\begin{split}
F(s,b) & =
\sqrt{8\pi}\,e^{-s^2/2}\,
\erf\!\Big(\frac{2b-s}{\sqrt2}\Big) \\
& +\frac{4}{s}\,e^{-2b^2}\Big(e^{(2b-s)s}-1\Big) \\
& +8\sqrt{\pi}\,e^{-b^2}\big[\erf(s-b)-\erf(b)\big] \\
& +\frac{2}{3}\,e^{-2b^2}\big(16b^3-12b^2 s+s^3+24b-12s\big).
\end{split}
\end{equation}
and support on $u \leq 2 v_e$. While we do not provide an explicit expression for the normalization, it can be found by imposing $\int f_\mathrm{rel}(u)du = 1.$

\section{Press-Schechter Halo Mass Functions on the \cmfast Grid}
\label{App:EPS}

Our modeling of energy injection from PBH accretion and \textit{p}-wave annihilation depends on calculations of the local halo mass function, which we detail here. Our treatment follows that of the EPS calculation performed internally by \cmfast, but we summarize the salient details here due to the importance of modeling the HMF for our projected sensitivities.

In order to determine the local halo mass function, we make use of that \dmcm has access to the discretized overdensity field $\delta(\mathbf{x}, z)$ on the lattice at a comoving resolution of $2\,\mathrm{Mpc}$ at all redshifts via \cmfast. To compute the conditional mass function, we follow the standard EPS formalism, see, \textit{e.g.}, \cite{2010gfe..book.....M} for more details. Throughout our EPS calculation, we make use of the spherical top-hat window function defined for a given radius $R$ in the spectral domain by
\begin{equation}
W_R(k) = 3 \left[\dfrac{\sin(k R)}{(k R)^3} - \dfrac{\cos(k R)}{(k R)^2} \right]
\label{eq:tophat}
\end{equation}
so that the halo mass $M$ is related to the comoving radius $R$ by
\begin{equation}
    M(R) = \frac{4 \pi }{3} (\Omega_\mathrm{DM} + \Omega_\mathrm{b}) \rho_c R^3
\end{equation}
where $\Omega_\mathrm{DM}$ and $\Omega_\mathrm{b}$ are the fraction abundances of DM and baryons at $z = 0$ and $\rho_c$ is the critical density at $z = 0$. Note that we are implicitly assuming that baryons and DM cluster identically. These cosmological parameters are taken to their Planck 2018 values \cite{Planck:2018vyg}.

First, because our overdensity field is specified at a comoving resolution of $2\,\mathrm{Mpc}$, we compute the density field smoothed at a radius of $R= 2\,\mathrm{Mpc}$ by convolving it with Eq.~\ref{eq:tophat}. We denote this smoothed field value $\delta_\mathrm{grid}(\mathbf{x}, z)$, and at each lattice point of the simulation, we calculate the local halo mass function $dN(\mathbf{x}, z |\delta_\mathrm{grid}(\mathbf{x}),z)/dM$ which depends on the redshift $z$ and the location through its conditioning on $\delta_\mathrm{grid}$.

To compute the conditional mass function, let $S(M)=\sigma^2(M)$ denote the variance of the smoothed density field on mass scale $M$, and define the corresponding values $S_1 = \sigma^2(M)$ and $S_\mathrm{grid} = \sigma^2(M_\mathrm{grid})$. We then define the linear overdensities as scaled by the growth factor
\begin{equation}
\delta_1(z)= \frac{\delta_c}{D(z)}, \qquad \delta_2(\mathbf{x}, z) = \frac{\delta_\mathrm{grid}(\mathbf{x}, z) }{D(z)},
\end{equation}
where \(\delta_c \approx 1.686\) is the critical collapse threshold. We define the conditional variable
\begin{equation}
\nu_{12}(\mathbf{x}, z)  \equiv \frac{\delta_1 - \delta_2(\mathbf{x}, z) }{\sqrt{S_1 - S_2}}.
\end{equation}
and the Press-Schechter multiplicity function
\begin{equation}
f_{\mathrm{PS}}(\nu_{12}) = \sqrt{\frac{2}{\pi}} \, \nu_{12} \, \exp\left(-\frac{\nu_{12}^2}{2}\right).
\end{equation}
so that the conditional mass function is then
\begin{equation}
\frac{\mathrm{d}N}{\mathrm{d}M}(M, \mathbf{x}, z) = \frac{(\Omega_\mathrm{DM}+\Omega_\mathrm{b}) \rho_0}{2 M^2} \left| \frac{\mathrm{d}S}{\mathrm{d}M} \right| \frac{ f_{\mathrm{PS}}(\nu_{12})}{ (S_1 - S_2)}.
\end{equation}
The unconditional Press-Schechter halo mass function, which is location-independent, is then trivially realized by taking $\delta_2 \rightarrow 0$ and $S_2 \rightarrow 0$.

For definiteness, we take the minimum halo mass to be $M_\mathrm{min} = 10^{-6}\,\mathrm{M}_\odot$, though this choice has negligible impact on our projected sensitivities as the signal is dominated by annihilation in larger halos.
Note that our conditional halo mass function computed via EPS is only defined for masses which are smaller than the total amount of matter in single lattice volume of our simulations, corresponding to $M_\mathrm{grid} \approx 3\times 10^{11} M_\odot$, meaning that our conditional mass function will never predict any abundance of halos $M_\mathrm{halo} \gtrsim 3\times 10^{11} \,\mathrm{M}_\odot$.  Accounting for the presence of halos larger than $10^{11}\,\mathrm{M}_\odot$ would require a detailed real-space treatment as they would be well-localized and resolved on the simulation lattice as compared to smaller halos, which can be treated with sub-grid via the EPS.\footnote{An analogous real-space treatment is indeed implemented  in \cmfast to account for the ionization from UV emission from large halos.}

However, as we argue, the effect of energy injection via either PBH accretion or \textit{p}-wave annihilations in these large halos is, in fact, subdominant in terms of driving \textit{sensitivity} (but not total energy injection) compared to accretion or annihilations in smaller halos. As we show in Fig.~\ref{fig:pwave-hmf-contribution}, at redshifts $z \gtrsim 10$, halos with masses greater than $3\times 10^{11} \,\mathrm{M}_\odot$ are sufficiently rare that they account for only a small fraction of the total annihilation or accretion in halos. At later redshifts with $z \lesssim 10$, standard astrophysical processes in those same very large halos dominate the total energy injection and even shut off the 21-cm signal as they fully ionize the IGM. As a result, relatively little sensitivity to exotic processes arises from late redshifts, and failing to resolve the contribution of \textit{p}-wave annihilation or PBH accretion in those late-forming large halos does not strongly affect our projected sensitivity. Moreover, accounting for their contributions would only serve to strengthen the projected sensitivity, so neglecting them is conservative.

\bibliography{main}
\end{document}